\documentclass[a4paper,11pt]{article}
\pdfoutput=1
\usepackage{jheppub}
\usepackage[T1]{fontenc}
\usepackage{needspace}
\usepackage{array,booktabs}
\usepackage{tikz}
\usetikzlibrary{decorations.pathmorphing}

\newcommand{\rmd}{\mathrm{d}}

\let\originaltableofcontents\tableofcontents
\renewcommand{\tableofcontents}{{\small\originaltableofcontents}}

\begin{document}
\title{Conservative tidal effects to third post-Minkowskian order in scalar-tensor theories}

\author{Zhengwen Liu}
\emailAdd{zhengwen.liu@seu.edu.cn}
\affiliation{School of Physics and Shing-Tung Yau Center, Southeast University, Nanjing 210096, China}

\author{and Hongbin Wang}
\emailAdd{101300488@seu.edu.cn}

\abstract{Using worldline effective field theory, we compute the conservative finite-size contributions to the impulse and scattering angle of nonspinning compact binaries in massless scalar-tensor theories with a scalar-Gauss-Bonnet coupling through third post-Minkowskian (3PM) order. We retain the full velocity dependence at linear order in the tidal response coefficients. The calculation covers the scalar dipole and quadrupole and the electric, magnetic, and mixed quadrupoles and octupoles. Among them, the scalar dipole enters the dynamics already at third post-Newtonian (3PN) order, two orders before the gravitational quadrupole. At 3PM, the tidal correction to the angle also depends on the first scalar derivatives of the responses and contains terms linear in the Gauss-Bonnet coupling. The results reproduce the general-relativistic tidal angle and the post-Newtonian tidal dynamics through 5PN, and a test-body calculation confirms their leading order in the mass ratio, including the Gauss-Bonnet terms for a companion with vanishing sensitivity.}
\maketitle

\section{Introduction}
\label{sec:introduction}

The direct detection of gravitational waves from coalescing compact binaries has opened a new avenue for testing gravity in the dynamical strong-field regime~\cite{LIGOScientific:2016lio,LIGOScientific:2021sio,Yunes:2024lzm}.
Distinguishing general relativity from well-motivated alternatives with these observations requires accurate predictions for compact-binary dynamics.
Scalar-tensor theories, a well-studied class of such alternatives, feature a long-range scalar field that contributes to interactions between compact bodies~\cite{Brans:1961sx,Damour:1992we,Julie:2024fwy}.
The sensitivity of each body characterizes its coupling to this field and measures how the body's mass depends on the external scalar~\cite{Eardley:1975fgi,Julie:2022huo}.
In some theories, spontaneous scalarization can enhance these couplings in neutron stars~\cite{Damour:1993hw,Doneva:2022ewd}, while a scalar-Gauss-Bonnet interaction can support black-hole scalar hair~\cite{Sotiriou:2013qea,Doneva:2017bvd,Silva:2017uqg}.

Beyond these couplings, finite-size effects depend on the tidal response of each body to the metric and scalar fields of its companion.
In general relativity, the leading conservative tidal effect enters at fifth post-Newtonian (5PN) order through the quadrupolar response~\cite{Flanagan:2007ix,Hinderer:2007mb}, which GW170817 has constrained~\cite{LIGOScientific:2017vwq}.
A long-range scalar field introduces additional responses, some of which enter at lower post-Newtonian order. In particular, a gradient of the external scalar induces a scalar dipole moment, which contributes from 3PN order, two orders before the gravitational quadrupole~\cite{Bernard:2019yfz}; black holes with scalar hair can also exhibit such dipoles~\cite{vanGemeren:2023rhh}.
Quadrupolar responses include scalar, metric, and mixed contributions~\cite{Creci:2023cfx,Bernard:2023eul}. Like the masses, these response coefficients depend on the external scalar field; their values have been computed for neutron stars in~\cite{Creci:2023cfx}.

The separation between the size of a compact object and the orbital length scale allows its internal response to be incorporated systematically into the two-body dynamics~\cite{Porto:2016pyg}.
Effective field theory builds on this scale separation: long-wavelength fields mediate the binary interaction, while additional couplings to the metric and scalar tidal fields describe the bodies' deformations~\cite{Goldberger:2004jt,Porto:2016zng}.
Matching to the compact object's response fixes the coefficients of these couplings~\cite{Kol:2011vg,Creci:2021rkz,Ivanov:2024sds}; see~\cite{Rodriguez:2026iot} for a recent review.
The same framework can also accommodate dynamical tides~\cite{Steinhoff:2016rfi}.

In the post-Newtonian regime, effective field theory has yielded the conservative dynamics in general relativity through 4PN order~\cite{Foffa:2019rdf,Foffa:2019yfl} and static interactions at higher orders~\cite{Foffa:2019hrb,Blumlein:2019zku,Liu:2026sib}. It also describes spin effects~\cite{Porto:2005ac,Porto:2010tr,Cho:2022syn}, as well as gravitational radiation and radiation reaction~\cite{Galley:2015kus,Amalberti:2024jaa,Porto:2026fsd}.
Tidal contributions to the dynamics and waveform have been computed to high post-Newtonian order~\cite{Henry:2019xhg,Henry:2020ski,Mandal:2023hqa}.
In scalar-tensor theories, results for the conservative dynamics without tidal effects extend through 3PN order~\cite{Bernard:2018hta,Bernard:2018ivi,Julie:2022qux}, and the scattering angle for hyperbolic encounters has also been computed~\cite{Jain:2023vlf}. Tensor and scalar waveforms are also available at the first post-Newtonian orders~\cite{Lang:2013fna,Shiralilou:2021mfl,Bernard:2022noq}.
Effective-field-theory calculations have extended these results~\cite{Kuntz:2019zef,Almeida:2024uph}, most recently determining the static part of the dynamics through 5PN order~\cite{Almeida:2026yvy}.
Results for the conservative tidal dynamics in these theories extend through next-to-next-to-leading post-Newtonian order~\cite{Bernard:2023eul}, and tidal contributions to fluxes and waveforms have also been calculated~\cite{Creci:2024wfu,Dones:2025zbs}.

Unlike the post-Newtonian expansion, the post-Minkowskian (PM) approach is perturbative in Newton's constant while retaining the exact dependence on the relative velocity~\cite{Bertotti:1956pxu,Westpfahl:1979gu,Damour:2016gwp}.
It therefore describes scattering directly, and its results can be mapped to bound orbits~\cite{Kalin:2019rwq,Kalin:2019inp}.
Scattering amplitudes~\cite{Cheung:2018wkq,Kosower:2018adc} and the effective field theory methods~\cite{Kalin:2020mvi,Mogull:2020sak} provide complementary routes to scattering observables. Both approaches have yielded the conservative dynamics at 3PM~\cite{Bern:2019nnu,Bern:2019crd,Brandhuber:2021eyq,Kalin:2020fhe}, as well as radiation-reaction effects at this order~\cite{DiVecchia:2020ymx,DiVecchia:2021bdo,Herrmann:2021tct,Bjerrum-Bohr:2021din,Kalin:2022hph} and spin effects at 2PM~\cite{Bern:2020buy,Kosmopoulos:2021zoq,Liu:2021zxr}.
At 4PM, the potential and tail contributions to the conservative dynamics were determined from amplitudes~\cite{Bern:2021dqo,Bern:2021yeh} and in effective field theory~\cite{Dlapa:2021npj,Dlapa:2021vgp}. The latter approach has also yielded the impulse including radiation reaction~\cite{Dlapa:2022lmu} and the local-in-time conservative dynamics~\cite{Dlapa:2024cje}.
At 5PM, conservative results are available at first self-force order~\cite{Driesse:2024xad}, including the local-in-time dynamics~\cite{Dlapa:2025biy}; nonlocal tail contributions have been obtained through tenth self-force order~\cite{Dlapa:2026oyq}.
On the computational side, graph-based algorithms have been developed to construct worldline effective-field-theory integrands~\cite{Kalin:2026txf}.

In the post-Minkowskian expansion, tidal effects were obtained at next-to-leading order in effective field theory~\cite{Kalin:2020lmz} and with scattering amplitudes~\cite{Cheung:2020sdj,Haddad:2020que,Bern:2020uwk}, alongside results from other approaches~\cite{Bini:2020flp}. Tidal calculations now extend through 4PM~\cite{Jakobsen:2023pvx}, and radiative tidal effects have also been computed~\cite{Mougiakakos:2022sic,Heissenberg:2022tsn}.

Beyond general relativity, post-Minkowskian calculations extend to higher-curvature corrections~\cite{Brandhuber:2019qpg,AccettulliHuber:2020oou} and to theories with additional light fields~\cite{Carrillo-Gonzalez:2021mqj}.
In massless scalar-tensor theories, the conservative scattering angle without tidal effects is known through 3PM~\cite{Bernard:2026old}. Effective field theory has also yielded both the impulse and the scattering angle through 3PM in the presence of a scalar coupling to the Gauss-Bonnet invariant~\cite{Almeida:2026abr}.
The general post-Minkowskian framework for multi-scalar theories developed in~\cite{Wilson-Gerow:2025xhr} includes the tidal terms known from general relativity.

In this paper, we compute the conservative finite-size contributions to the impulse and scattering angle of nonspinning compact binaries through 3PM and to linear order in the tidal response coefficients. Building on the 3PM results without tidal effects~\cite{Almeida:2026abr}, the calculation extends the general-relativistic tidal analysis~\cite{Kalin:2020lmz} to scalar-tensor theories.
The impulse is obtained by integrating the effective equations of motion along the perturbed worldlines, and the scattering angle then follows~\cite{Kalin:2020fhe,Almeida:2026abr}.
In addition to the electric and magnetic quadrupoles and octupoles in general relativity, our results include the scalar dipole and quadrupole and the mixed quadrupole and octupole. They also account for the scalar dependence of every response, whose first derivative enters the angle at 3PM.
The Gauss-Bonnet interaction also contributes to the tidal terms at this order through its lowest vertex.
All dependence on the bodies is encoded in their asymptotic masses, sensitivities, and dimensionless responses.

We validate the results in three limits, setting the Gauss-Bonnet coupling to zero unless stated otherwise.
First, for vanishing sensitivities, the scattering angle reproduces the general-relativistic tidal result through 3PM for the electric and magnetic quadrupoles and octupoles~\cite{Kalin:2020lmz,Cheung:2020sdj}.
Second, we consider a tidally deformed probe in the static scalar-tensor background of a point mass. This independent calculation checks the velocity dependence of the 2PM coefficients, the terms involving response derivatives, and the mass-ratio-independent parts of the remaining 3PM coefficients. For a companion with vanishing sensitivity, it also tests the Gauss-Bonnet terms.
Third, the low-velocity expansion of the angle agrees term by term with the finite-size contributions derived from the post-Newtonian conserved energy and angular momentum~\cite{Bernard:2023eul}, for all 2PM and 3PM terms through absolute 5PN order.

The paper is organized as follows.
Section~\ref{sec:action} introduces the bulk and worldline actions and their weak-field expansion.
Section~\ref{sec:pm} summarizes the effective-field-theory framework used to compute the scattering observables.
Section~\ref{sec:results} presents the impulse and scattering angle together with the consistency checks, and Section~\ref{sec:conclusions} concludes.
The appendices collect the Jordan-frame dictionary, the coefficient functions, and the details of the post-Newtonian comparison.

\section{Scalar-tensor theories and tidally deformed bodies}
\label{sec:action}

The two-body interaction in scalar-tensor theories depends on the bodies' scalar couplings and their tidal responses.
To describe both, we add tidal couplings to the action used in~\cite{Almeida:2026abr} and follow its conventions.

\subsection{The bulk action}
\label{sec:bulk}

The metric and a massless scalar mediate the gravitational interaction, with an additional scalar coupling to the Gauss-Bonnet invariant.
In the Einstein frame, the bulk action reads~\cite{Almeida:2026abr}
\begin{align}
 S_{\mathrm{g}}&=-\frac{2}{\kappa^2}\int\rmd^4x\sqrt{-g}
 \left[R-2g^{\mu\nu}
 (\partial_\mu\varphi)(\partial_\nu\varphi)-\alpha f(\varphi)\mathcal G\right],
 \label{eq:Einstein-action}
\end{align}
with $\eta_{\mu\nu}=\mathrm{diag}(1,-1,-1,-1)$, $\kappa^2=32\pi G$, $c=1$, and
\begin{align}
 \mathcal G&=R^2-4R_{\mu\nu}R^{\mu\nu}
 +R_{\mu\nu\rho\sigma}R^{\mu\nu\rho\sigma}.
 \label{eq:GB-invariant}
\end{align}
The coupling $\alpha$ has dimensions of length squared. Since the post-Newtonian results used in our comparison are formulated in the Jordan frame, Appendix~\ref{app:laura-notation} gives the corresponding dictionary.

\subsection{The worldline action}
\label{sec:worldline}
At separations large compared with their sizes, compact bodies in scalar-tensor theories can be represented by point particles with scalar-dependent masses $M_a(\varphi)$~\cite{Eardley:1975fgi,Damour:1992we}. Following~\cite{Almeida:2026abr}, the point-particle action is
\begin{align}
 S_\mathit{pp}&=-\sum_{a=1}^2\int\rmd\tau_a\,M_a(\varphi).
 \label{eq:Spp}
 \end{align}
These effective masses encode each body's coupling to the external scalar field and already include its response to a constant scalar background. Adiabatic tidal deformations add terms quadratic in the tidal fields, which describe the bodies' linear response~\cite{Bini:2012gu,Kalin:2020lmz}; in scalar-tensor theories, they include scalar and mixed scalar-tensor contributions~\cite{Bernard:2019yfz,Creci:2023cfx,Bernard:2023eul}. The full worldline action is
\begin{align}
 S_{\mathrm{WL}}&=S_\mathit{pp}+S_\mathit{fs}^\text{(s)}+S_\mathit{fs}^\text{(g)}+S_\mathit{fs}^\text{(gs)},
 \label{eq:WL-decomposition}
\end{align}
where the finite-size terms describe scalar-induced, metric, and mixed tidal effects, respectively:
\begin{align}
 S_\mathit{fs}^\text{(s)} &=-\sum_{a=1}^2\int\rmd\tau_a
 \sum_{\ell=1}^{\infty}\frac{\lambda_a^{(\ell)}(\varphi)}{2\ell!}
 (\nabla_L^\perp\varphi)_a(\nabla_\perp^L\varphi)_a,
 \label{eq:Sscalar}
\\[0.5 em]
 S_\mathit{fs}^\text{(g)} &=\sum_{a=1}^2\int\rmd\tau_a
 \sum_{\ell=2}^{\infty}\frac{1}{2\ell!}
 \left[\rho_a^{(\ell)}(\varphi)G_L^aG_a^L
 +\frac{\ell}{\ell+1}\xi_a^{(\ell)}(\varphi)H_L^aH_a^L\right],
 \label{eq:Sgravity}
\\[0.5 em]
 S_\mathit{fs}^\text{(gs)} &=-\sum_{a=1}^2\int\rmd\tau_a
 \sum_{\ell=2}^{\infty}\frac{\zeta_a^{(\ell)}(\varphi)}{\ell!}
 G_a^L(\nabla_L^\perp\varphi)_a.
 \label{eq:Smixed}
\end{align}
Here $L=\mu_1\cdots\mu_\ell$ is a multi-index, and all fields are evaluated on the corresponding worldline. The electric and magnetic tidal moments are defined by
\begin{align}
 G^a_{\mu_1\cdots\mu_\ell} &=-
 \left[\nabla^\perp_{\langle\mu_1}\cdots
 \nabla^\perp_{\mu_{\ell-2}}
 C_{\mu_{\ell-1}\underline{\rho}\mu_\ell\rangle\sigma}\right]
 \dot{x}_a^\rho\dot{x}_a^\sigma,
 \label{eq:Gmoment}
\\[0.5 em]
 H^a_{\mu_1\cdots\mu_\ell} &=2
 \left[\nabla^\perp_{\langle\mu_1}\cdots
 \nabla^\perp_{\mu_{\ell-2}}
 C^*_{\mu_{\ell-1}\underline{\rho}\mu_\ell\rangle\sigma}\right]
 \dot{x}_a^\rho\dot{x}_a^\sigma.
 \label{eq:Hmoment}
\end{align}
Here $\dot{x}_a^\mu\equiv\rmd x_a^\mu(\tau_a)/\rmd\tau_a$ is the proper-time four-velocity, satisfying $g_{\mu\nu}\dot{x}_a^\mu\dot{x}_a^\nu=1$. The dual of the Weyl tensor $C_{\mu\nu\rho\sigma}$ is
$C^*_{\mu\rho\nu\sigma}
=\tfrac12\epsilon_{\mu\rho}{}^{\alpha\beta}C_{\alpha\beta\nu\sigma}$,
with $\epsilon_{0123}=+\sqrt{-g}$. Angle brackets denote symmetrization over the enclosed indices, excluding the underlined index, followed by trace removal with respect to the rest-space metric $g_{\mu\nu}-\dot{x}_{a\mu}\dot{x}_{a\nu}$. The projected derivatives, which also appear in the scalar and mixed operators, are
\begin{align}
 (\nabla_\mu^\perp)_a&=
 (\delta_\mu^\nu-\dot{x}_{a\mu}\dot{x}_a^\nu)\nabla_\nu,
 \qquad
 \nabla_L^\perp=\nabla_{\mu_1}^\perp\cdots\nabla_{\mu_\ell}^\perp.
 \label{eq:projected-derivative}
\end{align}
The order of differentiation is part of the operator definition: in a nested scalar derivative, each outer covariant derivative acts on both the inner projectors and the scalar. We adopt a Cartesian prescription in which external spacetime derivatives are taken with the contravariant components of the worldline velocity held fixed.

We retain the two lowest multipoles of each operator type, as summarized in Table~\ref{tab:operators}. Unlike the tidal moments $G_L$ and $H_L$, the scalar operators are not projected onto symmetric trace-free form. Consequently, matching their trace and nonlinear-derivative terms to a compact object requires information beyond static Love parameters. Similarly, the first scalar derivatives of the responses, which enter at 3PM, describe how each response varies with the background scalar and are not fixed by its value at a single background.

\begin{table}[t]
\centering
\begin{tabular}{@{}lllll@{}}
\toprule
Operator & Response & $\ell$ & Type & First contribution\\
\midrule
$(\nabla_L^\perp\varphi)_a(\nabla_\perp^L\varphi)_a$ & $\lambda_a^{(\ell)}$ & 1, 2 & scalar-induced & 2PM; 3PN, 5PN\\
$G_L^aG_a^L$ & $\rho_a^{(\ell)}$ & 2, 3 & electric & 2PM; 5PN, 7PN\\
$H_L^aH_a^L$ & $\xi_a^{(\ell)}$ & 2, 3 & magnetic & 2PM; 6PN, 8PN\\
$G_a^L(\nabla_L^\perp\varphi)_a$ & $\zeta_a^{(\ell)}$ & 2, 3 & mixed & 2PM; 5PN, 7PN\\
\bottomrule
\end{tabular}
\caption{Finite-size operators, the multipoles $\ell$ included in the results, and the PM and PN orders at which the operators first contribute. In each row, the two PN orders correspond, respectively, to the two values of $\ell$.}
\label{tab:operators}
\end{table}

\subsection{Weak-field expansion}
\label{sec:scalar-dependence}

We study weak-field encounters without assuming slow motion. The metric is expanded about Minkowski spacetime and the scalar about its constant asymptotic value:
\begin{align}
 g_{\mu\nu}&=\eta_{\mu\nu}+\kappa h_{\mu\nu},\qquad\varphi=\varphi_0+\kappa\phi.
 \label{eq:scalar-expansion}
\end{align}
The Gauss-Bonnet coupling function, the masses, and the response coefficients all depend on the scalar field, and we expand each of them about $\varphi_0$. For the coupling function,
\begin{align}
 f(\varphi) &=f_1(\kappa\phi)+\frac{f_2}{2}(\kappa\phi)^2
 +\frac{f_3}{6}(\kappa\phi)^3+\mathcal{O}(\kappa^4).
 \label{eq:f-expansion}
\end{align}
The coupling coefficients $f_k$ specify the theory and are not restricted to a particular model.
The expansion of the masses determines the scalar couplings of the bodies~\cite{Almeida:2026abr},
\begin{align}
 M_a(\varphi)&=m_a\Bigl[1+s_a(\kappa\phi)
 +\frac{s_a^2+s'_a}{2}(\kappa\phi)^2
 +\frac{s_a^3+3s_as'_a+s''_a}{6}(\kappa\phi)^3
 +\mathcal{O}(\kappa^4)\Bigr],
 \label{eq:mass-expansion}
\end{align}
where $m_a=M_a(\varphi_0)$ are the asymptotic masses. The sensitivities quantify the scalar dependence of the masses through logarithmic derivatives evaluated at $\varphi_0$:
\begin{align}
 s_a&=\left.\frac{\rmd\ln M_a}{\rmd\varphi}\right|_{\varphi_0},
 \qquad s'_a=\left.\frac{\rmd^2\ln M_a}{\rmd\varphi^2}\right|_{\varphi_0},
 \qquad s''_a=\left.\frac{\rmd^3\ln M_a}{\rmd\varphi^3}\right|_{\varphi_0}.
 \label{eq:mass-sensitivities}
\end{align}
The total mass, reduced mass, and symmetric mass ratio are
\begin{align}
 m&=m_1+m_2,
 \qquad \mu=\frac{m_1m_2}{m},
 \qquad \nu=\frac\mu m.
 \label{eq:mass-combinations}
\end{align}

We expand the response coefficients $F\in\{\lambda,\rho,\xi,\zeta\}$ in the same way,
\begin{align}
 F_a^{(\ell)}(\varphi)&=\sum_{k\geq0}\frac{F_{a,k}^{(\ell)}}{k!}(\kappa\phi)^{k},
 \qquad
 F_{a,k}^{(\ell)}=\left.\frac{\rmd^{k}F_a^{(\ell)}}{\rmd\varphi^{k}}\right|_{\varphi_0}.
 \label{eq:Taylor-series}
\end{align}
Here $a$ labels the body, $\ell$ the multipole, and $k$ the scalar-derivative order. Throughout, we work to linear order in the response coefficients.

To factor out each body's mass and gravitational length scale, we define the dimensionless responses
\begin{align}
 \bar{F}_{a,k}^{(\ell)}&=\frac{F_{a,k}^{(\ell)}}{m_a(Gm_a)^{2\ell}}.
 \label{eq:dimensionless-responses}
\end{align}
The normalization uses the constant asymptotic masses and therefore commutes with scalar differentiation.

The bodies may be strongly self-gravitating even though their mutual interaction remains weak. The expansion requires small deflections with controlled higher-PM corrections, body sizes small compared with the rest-frame variation scale of the external fields, and tidal responses that remain adiabatic and perturbative. The mass and response expansions must also remain controlled over the scalar excursion induced by the companion.

\section{Effective field theory}
\label{sec:pm}

During a scattering encounter, the metric and scalar interactions deflect the bodies away from straight-line trajectories, while tidal deformations modify the momentum exchanged between them. Worldline effective field theory incorporates these finite-size effects through the tidal response coefficients introduced above~\cite{Kalin:2020mvi}.

Integrating out the metric and scalar perturbations yields an effective action for the trajectories,
\begin{align}
 e^{iS_{\rm eff}[x_a]}&=\int\mathcal Dh\,\mathcal D\phi\;e^{iS_{\mathrm{g}}[h,\phi]+iS_{\rm gf}[h]+iS_{\mathrm{WL}}[x_a,h,\phi]},
 \label{eq:EFT}
\end{align}
where $S_{\rm gf}$ is the gauge-fixing term. At the classical saddle point, the fields satisfy their equations of motion with the two worldlines as sources. Integrating the resulting worldline equations of motion along the perturbed trajectories yields the impulse and hence the scattering angle. This procedure follows the calculations without tidal effects~\cite{Kalin:2020fhe,Almeida:2026abr} and the general-relativistic tidal calculation~\cite{Kalin:2020lmz}.

\subsection{Equations of motion}
\label{sec:dynamics}

Because each body's equations of motion depend on both trajectories, the two deflections must be determined together order by order.
We choose a worldline parameter for which the free action is quadratic in the velocity~\cite{Kalin:2020mvi,Almeida:2026abr} and write
\begin{align}
 S_{\rm eff}&=-\sum_a\frac{m_a}{2}\int\rmd\tau_a\,\eta_{\mu\nu}\dot x_a^\mu\dot x_a^\nu+S_{\rm int}[x_1,x_2],
 \label{eq:Seff-split}
\end{align}
where dots denote derivatives with respect to $\tau_a$ and $S_{\rm int}$ collects all interactions.
This parameter coincides with proper time in the asymptotic past and future.
Varying the action yields the equations of motion,
\begin{align}
 m_a\,\ddot x_a^\mu(\tau_a)&=-\eta^{\mu\nu}\,\frac{\delta S_{\rm int}[x_1,x_2]}{\delta x_a^\nu(\tau_a)}.
 \label{eq:eom}
\end{align}
At linear order in the response coefficients, $S_{\rm int}$ contains the point-particle interactions and terms with a single tidal-response insertion.

We select the scattering solution by imposing $\dot x_a^\mu\to u_a^\mu$ as $\tau_a\to-\infty$, with $u_a^2=1$. The corresponding unperturbed trajectories are $x_{a,0}^\mu=b_a^\mu+u_a^\mu\tau_a$. The incoming velocities determine the Lorentz factor $\gamma$ and the relative speed $v$,
\begin{align}
 \gamma&=u_1\cdot u_2=\frac1{\sqrt{1-v^2}},\qquad \gamma>1,
 \label{eq:gamma-def}
\end{align}
and $b^\mu=b_1^\mu-b_2^\mu$ is the impact parameter, with $b\cdot u_1=b\cdot u_2=0$, $b^2=-b_e^2$, and $\hat b^\mu\equiv b^\mu/b_e$.

We solve the equations of motion perturbatively about these straight-line trajectories. The perturbed worldlines and their incoming conditions are
\begin{align}
 x_a^\mu(\tau_a)&=b_a^\mu+u_a^\mu\tau_a+\sum_{n\geq1}\delta^{(n)}x_a^\mu(\tau_a),
 \qquad \frac{\rmd}{\rmd\tau_a}\delta^{(n)}x_a^\mu\to0\quad\text{for}\quad\tau_a\to-\infty,
 \label{eq:trajectory-expansion}
\end{align}
with $\delta^{(n)}x_a^\mu$ of order $G^n$.
At each order, we evaluate the right-hand side of~\eqref{eq:eom} on the lower-order trajectories and integrate twice, imposing the incoming boundary conditions.
In four dimensions, the long-range point-particle interaction produces a logarithmic longitudinal displacement; the corresponding integrals are evaluated using the incoming prescription introduced in~\cite{Kalin:2020mvi}.
The 1PM deflections due to single-graviton or single-scalar exchange are given in~\cite{Almeida:2026abr} and, for general relativity, in~\cite{Kalin:2020mvi}.

\subsection{The impulse}
\label{sec:eft-impulse}

Integrating the equations of motion over the encounter yields the impulse, defined as the change in each body's momentum between the asymptotic incoming and outgoing states,
\begin{align}
 \Delta p_a^\mu&\equiv m_a\int_{-\infty}^{+\infty}\rmd\tau_a\,\ddot x_a^\mu
 =-\eta^{\mu\nu}\int_{-\infty}^{+\infty}\rmd\tau_a\,\frac{\delta S_{\rm int}[x_1,x_2]}{\delta x_a^\nu(\tau_a)},
 \label{eq:impulse-def}
\end{align}
with the right-hand side evaluated along the perturbed trajectories.

\subsection{Finite-size vertices and power counting}
\label{sec:vertices}

\begin{figure}[t]
\centering
\begin{tikzpicture}[black,line width=0.7pt,
 grav/.style={decorate,decoration={snake,amplitude=1.3pt,segment length=4.6pt}},
 scal/.style={dashed},
 vertex/.style={circle,fill=black,inner sep=0pt,minimum size=5.5pt},
 lab/.style={font=\footnotesize,align=center,anchor=north,inner sep=1.5pt},
 cpl/.style={font=\footnotesize,anchor=north,inner sep=1.5pt,text height=2.1ex,text depth=0.8ex}]
\begin{scope}[xshift=0.0cm,yshift=0cm]
 \draw[scal] (0,0)--(0,1.12);
 \node[vertex] at (0,0) {};
 \node[cpl] (t) at (0,-0.16) {$m_as_a$};
 \node[lab] at (t.south) {(a)};
\end{scope}
\begin{scope}[xshift=2.3cm,yshift=0cm]
 \draw[grav] (0,0)--(0,1.12);
 \node[vertex] at (0,0) {};
 \node[cpl] (t) at (0,-0.16) {$m_a$};
 \node[lab] at (t.south) {(b)};
\end{scope}
\begin{scope}[xshift=0.0cm,yshift=-2.7cm]
 \draw[scal] (0,0)--(-0.42,1.05); \draw[scal] (0,0)--(0.42,1.05);
 \node[vertex] at (0,0) {};
 \node[cpl] (t) at (0,-0.16) {$m_a(s_a^2+s'_a)$};
 \node[lab] at (t.south) {(c)};
\end{scope}
\begin{scope}[xshift=2.3cm,yshift=-2.7cm]
 \draw[grav] (0,0)--(-0.42,1.05); \draw[scal] (0,0)--(0.42,1.05);
 \node[vertex] at (0,0) {};
 \node[cpl] (t) at (0,-0.16) {$m_as_a$};
 \node[lab] at (t.south) {(d)};
\end{scope}
\begin{scope}[xshift=6.15cm,yshift=0cm]
 \draw[scal] (0,0)--(-0.42,1.05); \draw[scal] (0,0)--(0.42,1.05);
 \node[vertex] at (0,0) {};
 \node[cpl] (t) at (0,-0.16) {$\lambda_{a,0}^{(\ell)}$};
 \node[lab] at (t.south) {(e)};
\end{scope}
\begin{scope}[xshift=8.45cm,yshift=0cm]
 \draw[grav] (0,0)--(-0.42,1.05); \draw[scal] (0,0)--(0.42,1.05);
 \node[vertex] at (0,0) {};
 \node[cpl] (t) at (0,-0.16) {$\zeta_{a,0}^{(\ell)}$};
 \node[lab] at (t.south) {(f)};
\end{scope}
\begin{scope}[xshift=10.75cm,yshift=0cm]
 \draw[grav] (0,0)--(-0.42,1.05); \draw[grav] (0,0)--(0.42,1.05);
 \node[vertex] at (0,0) {};
 \node[cpl] (t) at (0,-0.16) {$\rho_{a,0}^{(\ell)}$, $\xi_{a,0}^{(\ell)}$};
 \node[lab] at (t.south) {(g)};
\end{scope}
\begin{scope}[xshift=5.0cm,yshift=-2.7cm]
 \draw[scal] (0,0)--(-0.45,1.0); \draw[scal] (0,0)--(0.45,1.0); \draw[scal] (0,0)--(0,1.12);
 \node[vertex] at (0,0) {};
 \node[cpl] (t) at (0,-0.16) {$\lambda_{a,1}^{(\ell)}$};
 \node[lab] at (t.south) {(h)};
\end{scope}
\begin{scope}[xshift=7.3cm,yshift=-2.7cm]
 \draw[scal] (0,0)--(-0.45,1.0); \draw[scal] (0,0)--(0.45,1.0); \draw[grav] (0,0)--(0,1.12);
 \node[vertex] at (0,0) {};
 \node[cpl] (t) at (0,-0.16) {$\lambda_{a,0}^{(\ell)}$, $\zeta_{a,1}^{(\ell)}$};
 \node[lab] at (t.south) {(i)};
\end{scope}
\begin{scope}[xshift=9.6cm,yshift=-2.7cm]
 \draw[grav] (0,0)--(-0.45,1.0); \draw[grav] (0,0)--(0.45,1.0); \draw[scal] (0,0)--(0,1.12);
 \node[vertex] at (0,0) {};
 \node[cpl] (t) at (0,-0.16) {$\rho_{a,1}^{(\ell)}$, $\xi_{a,1}^{(\ell)}$, $\zeta_{a,0}^{(\ell)}$};
 \node[lab] at (t.south) {(j)};
\end{scope}
\begin{scope}[xshift=11.9cm,yshift=-2.7cm]
 \draw[grav] (0,0)--(-0.45,1.0); \draw[grav] (0,0)--(0.45,1.0); \draw[grav] (0,0)--(0,1.12);
 \node[vertex] at (0,0) {};
 \node[cpl] (t) at (0,-0.16) {$\rho_{a,0}^{(\ell)}$, $\xi_{a,0}^{(\ell)}$};
 \node[lab] at (t.south) {(k)};
\end{scope}
\end{tikzpicture}
\caption{Worldline vertices for tidal corrections to the impulse through 3PM. Dots denote vertices, wavy lines gravitons, and dashed lines scalar fluctuations. (a)--(d)~Point-particle vertices with one and two fields. (e)--(g)~Two-field tidal vertices. \mbox{(h)--(k)}~Three-field tidal vertices; response derivatives add a scalar leg, while metric expansions add a graviton leg.}
\label{fig:vertices}
\end{figure}
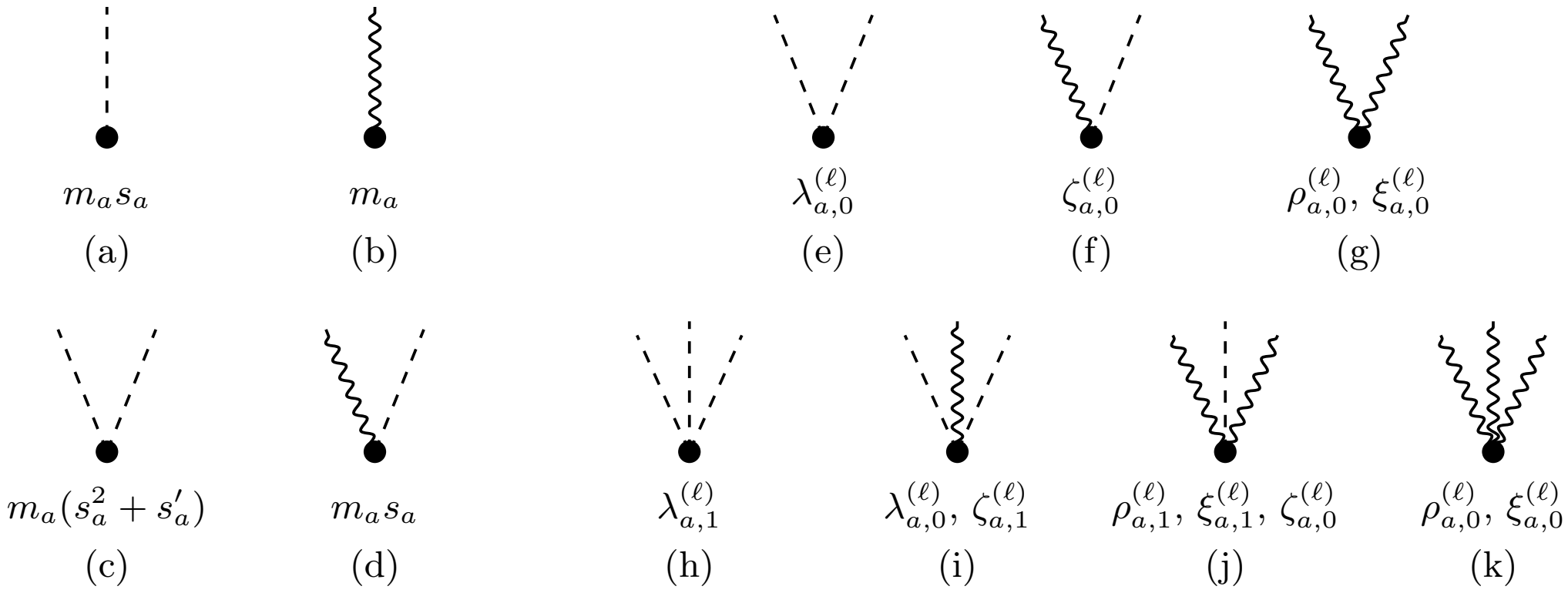

The tidal corrections involve two types of worldline vertices, point-particle and finite-size ones. The point-particle vertices follow from the unit-einbein form of \eqref{eq:Spp}~\cite{Kalin:2020mvi,Almeida:2026abr},
\begin{align*}
 S_{pp}=-\frac12\sum_a\int\rmd\tau_a\,M_a(\varphi)
 \bigl(g_{\mu\nu}\dot x_a^\mu\dot x_a^\nu+1\bigr).
\end{align*}
This action is linear in $h_{\mu\nu}$, so a two-leg point-particle vertex contains either two scalars or one scalar and one graviton; Figure~\ref{fig:vertices}(a)--(d) shows the one- and two-leg vertices together with their mass couplings. Point-particle vertices with three or more legs first enter the tidal corrections at 4PM.

Tidal vertices couple to at least two fields, so finite-size effects first enter at 2PM.
Expanding the tidal operators about flat space and the asymptotic scalar gives the finite-size vertices; the bulk vertices are given in~\cite{Almeida:2026abr}.
At Taylor order $k=0$, each operator starts at quadratic order in the field fluctuations: the scalar operator couples to two scalar fluctuations, the mixed operator to one graviton and one scalar fluctuation, and the electric and magnetic operators to two gravitons (Figure~\ref{fig:vertices}(e)--(g)).
Each additional field in a worldline vertex contributes one power of $\kappa$; the PM order is determined by the complete diagram.
At leading order, the two field lines from a tidal vertex connect to the companion, giving a contribution of order $G^2$. Throughout, $n$PM denotes order $G^n$ at fixed dimensionful response coefficients, as given by this counting. In terms of the dimensionless responses~\eqref{eq:dimensionless-responses}, an $n$PM term carries an additional factor of order $(Gm/b_e)^{2\ell}$, and post-Newtonian orders are counted with $\bar F_{a,k}^{(\ell)}=\mathcal O(1)$.

Vertices with additional field legs arise from the scalar dependence of the responses and the metric dependence of the operators. The $k=1$ Taylor term adds a scalar leg and introduces the response derivative $F_{a,1}^{(\ell)}$, while expanding the operators in the metric adds a graviton leg at $k=0$. These three-field vertices first contribute at 3PM. Figure~\ref{fig:vertices}(h)--(k) shows the four resulting field combinations. Further expansion generates vertices with more field legs, which contribute only beyond 3PM.

Because $\mathcal G$ is quadratic in the curvature, the lowest Gauss-Bonnet bulk vertex couples one scalar fluctuation to two gravitons. This vertex is proportional to $\alpha f_1$ and has two more derivatives than the cubic Einstein--Hilbert and scalar-kinetic vertices, producing the factor $\alpha f_1/b_e^2$ relative to the other 3PM terms in the angle.
In the sector with one tidal insertion, the term proportional to $f_2$ contains at least two scalar fluctuations and contributes only from 4PM on.

\subsection{From the impulse to the scattering angle}
\label{sec:angle}

For elastic conservative scattering, the magnitude of the momentum transfer determines the scattering angle,
\begin{align}
 \chi&=2\arcsin\left(
 \frac{\sqrt{-(\Delta p_1)^2}}{2p_\infty}\right),
 \label{eq:angle-from-impulse}
\end{align}
where $p_\infty$ is the magnitude of the incoming momentum in the center-of-mass frame. In terms of the total energy $E$ in this frame, the energy ratio $\Gamma$ and the momentum $p_\infty$ are
\begin{align}
 \Gamma&=\frac{E}m
 =\sqrt{1+2\nu(\gamma-1)},
%
\qquad
 p_\infty=\frac\mu\Gamma\sqrt{\gamma^2-1}.
 \label{eq:asymptotic-kinematics}
\end{align}

\section{Impulse and scattering angle}
\label{sec:results}

\begin{figure}[t]
\centering
\includegraphics[width=\linewidth]{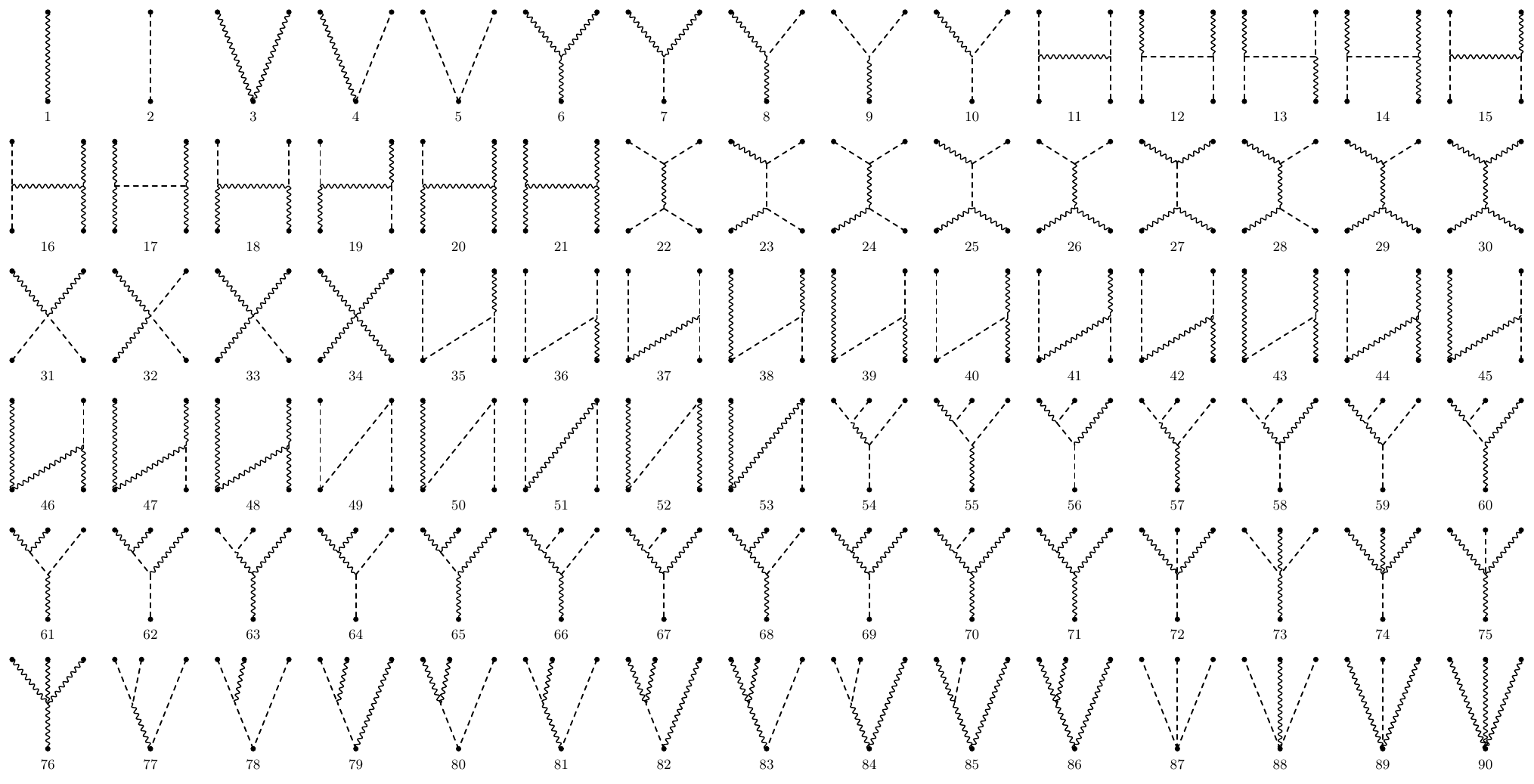}%
\caption{Point-particle and tidal diagram topologies through 3PM. Wavy and dashed lines denote graviton and scalar propagators, respectively. Filled dots mark worldline vertices, while junctions without dots mark bulk vertices. Crossings without junctions are not vertices.}
\label{fig:feynman-diagrams}
\end{figure}

The 3PM tidal correction to the impulse combines direct contributions evaluated on straight trajectories with iteration terms generated by lower-order perturbations of the two worldlines. Figure~\ref{fig:feynman-diagrams} displays the point-particle and tidal topologies through 3PM. The leading tidal exchange (diagrams 3--5) receives 3PM corrections from bulk interactions, two-leg point-particle vertices on the companion, and three-leg tidal vertices (diagrams 87--90). In addition, the 1PM point-particle deflections modify the leading tidal interaction, while the 2PM tidal deflections modify the 1PM point-particle interaction~\cite{Kalin:2020fhe,Kalin:2020lmz,Almeida:2026abr}. These iterated contributions incorporate both the position and velocity dependence of the worldline couplings.

Through 3PM, tidal insertions change the momentum numerators but leave the propagator structure unchanged, so these contributions involve the same integral families as the point-particle calculation~\cite{Kalin:2020fhe,Kalin:2020lmz}. Integration over the worldline parameters leaves one- and two-loop integrals at 2PM and 3PM, respectively. These integrals depend on the momentum transfer $q^\mu$, with $q\cdot u_1=q\cdot u_2=0$, and contain massless quadratic propagators, linear worldline denominators, and straight-worldline delta constraints. We use Feynman propagators with denominators $k^2+i0$ for both massless fields, while the incoming conditions fix the linear-pole prescriptions.

Integration-by-parts identities reduce these integrals to linear combinations of master integrals. Dimensional analysis fixes their $q^2$ scaling, while differential equations determine the $\gamma$ dependence from boundary data near $\gamma\to1$, evaluated by the method of regions~\cite{Kalin:2020fhe,Dlapa:2023hsl}. The two-loop basis and differential equations needed for the 3PM conservative calculation are given in~\cite{Dlapa:2023hsl}. Retaining only the potential region in these boundary data yields the conservative impulse. Finally, we sum the contributions with a tidal vertex on either body and Fourier transform the result with respect to $q^\mu$ to obtain the correction to the impulse in impact-parameter space.

\subsection{Impulse}
\label{sec:impulse}

The finite-size contribution to the impulse, linear in the response coefficients, decomposes into a transverse component along $\hat b^\mu$ and a longitudinal component along $w^\mu$, a linear combination of the incoming velocities:
\begin{align}
 \Delta p_1^\mu\big|_\text{tidal}&=-\Delta p_2^\mu\big|_\text{tidal}
 \nonumber\\
 &=\nu\sum_{F,\ell}\left(\frac{Gm}{b_e}\right)^{2\ell}\Biggl[\,\sum_{n=2}^{3}\left(\frac{Gm}{b_e}\right)^n
 \bigl(mc_{n,F}^{(\ell)}\,\hat b^\mu+d_{n,F}^{(\ell)}\,w^\mu\bigr)+\mathcal{O}(G^4)\Biggr],
 \label{eq:tidal-impulse-result}
\end{align}
where the sum runs over $F\in\{\lambda,\rho,\xi,\zeta\}$, with $\ell=1,2$ for $F=\lambda$ and $\ell=2,3$ otherwise, and
\begin{align*}
 w^\mu\equiv\frac{(\gamma m_1+m_2)u_1^\mu-(\gamma m_2+m_1)u_2^\mu}
                    {\sqrt{\gamma^2-1}}.
\end{align*}
In terms of the dimensionless responses, the transverse coefficients are
\begin{align}
 c_{2,F}^{(\ell)}
 &=-\nu\sqrt{\gamma^2-1}\Biggl[
 \left(\frac{m_1}{m}\right)^{2\ell-1}
 \bar{F}_{1,0}^{(\ell)}\mathcal A_F^{(\ell)}
 +(1\leftrightarrow2)\Biggr],\nonumber\\
 c_{3,F}^{(\ell)}
 &=-\nu^2\sqrt{\gamma^2-1}\Biggl[
 \left(\frac{m_1}{m}\right)^{2\ell-2}
 \Bigl(\bar{F}_{1,0}^{(\ell)}\mathcal B_F^{(\ell)}
       +\bar{F}_{1,1}^{(\ell)}\mathcal C_F^{(\ell)}\Bigr)
 +(1\leftrightarrow2)\Biggr].
 \label{eq:tidal-impulse-projection}
\end{align}
The coefficient functions $\mathcal A_F^{(\ell)}$, $\mathcal B_F^{(\ell)}$, and $\mathcal C_F^{(\ell)}$ are dimensionless and given explicitly in Appendix~\ref{app:tidal-coefficients} (computer-readable files are available in~\cite{STtidal:ancillary}).
The exchange $1\leftrightarrow2$ acts on the masses, sensitivities, and responses, including the body labels implicit in these functions; it leaves $c_{n,F}^{(\ell)}$ invariant and reverses both basis vectors.
The longitudinal coefficients are
\begin{align}
 d_{2,F}^{(\ell)}&=0,\qquad d_{3,F}^{(\ell)}=\frac{2\vartheta\,c_{2,F}^{(\ell)}}{\gamma^2-1},
 \label{eq:tidal-impulse-velocity-coefficients}
\end{align}
where $\vartheta\equiv2\gamma^2-1+s_1s_2$. Together with the point-particle contribution, the impulse satisfies the on-shell condition $(m_au_a+\Delta p_a)^2=m_a^2$ through 3PM at linear order in the response coefficients.

At 2PM, the coefficients $\mathcal A_F^{(\ell)}$ are $\pi$ times rational functions of $\gamma$ and multiply the response values at $\varphi_0$. The scalar and mixed coefficients are proportional to $s_2^2$ and $s_2$, respectively, while the electric and magnetic ones are independent of the sensitivities.

At 3PM, $\mathcal B_F^{(\ell)}$ multiplies $\bar{F}_{1,0}^{(\ell)}$, while $\mathcal C_F^{(\ell)}$ multiplies its first scalar derivative $\bar{F}_{1,1}^{(\ell)}$. The functions $\mathcal C_F^{(\ell)}$ are rational in $\gamma$ and proportional to $s_2^3$ in the scalar sector, to $s_2$ in the electric and magnetic sectors, and to $s_2^2$ in the mixed sector. We organize each $\mathcal B_F^{(\ell)}$ by the mass ratio, as a term independent of $m_1/m_2$ plus a term proportional to $m_1/m_2$; both terms contain explicit Gauss-Bonnet corrections proportional to $\alpha f_1/b_e^2$. The derivative $s'_2$ of the companion's sensitivity appears only in the terms proportional to $m_1/m_2$, and only in the scalar and mixed sectors.

Logarithms first appear at 3PM, in $\mathcal B_F^{(\ell)}$, and can be expressed in terms of the velocity variable
\begin{align}
 x&=\gamma-\sqrt{\gamma^2-1},\qquad
 \log x=-\operatorname{arccosh}\gamma.
 \label{eq:tidal-x-variable}
\end{align}
In this organization, the terms independent of $m_1/m_2$ are rational functions of $\gamma$, while the terms proportional to $m_1/m_2$ also contain rational functions multiplying $\log x/\sqrt{\gamma^2-1}$, and the same holds for the Gauss-Bonnet contribution.

\Needspace{9\baselineskip}
\subsection{Scattering angle}
\label{sec:angle-result}

Expanding~\eqref{eq:angle-from-impulse} to linear order in the response coefficients yields the tidal correction $\Delta\chi$ to the scattering angle,
\begin{align}
 \frac{\Delta\chi}{\Gamma}&=-\frac{\vartheta}{|\vartheta|\sqrt{\gamma^2-1}}\sum_{F,\ell}\left(\frac{Gm}{b_e}\right)^{2\ell}\Biggl[
 \left(\frac{Gm}{b_e}\right)^2c_{2,F}^{(\ell)}
 +\left(\frac{Gm}{b_e}\right)^3c_{3,F}^{(\ell)}+\mathcal{O}(G^4)\Biggr].
 \label{eq:tidal-angle-result}
\end{align}
Through 3PM, only the transverse coefficients contribute, and the factor $\vartheta/|\vartheta|$ arises from the 1PM point-particle impulse, whose transverse component is proportional to $-\vartheta\,\hat b^\mu$~\cite{Almeida:2026abr}.

\subsection{Consistency checks}
\label{sec:checks}
The general relativistic, test-body, and post-Newtonian limits provide complementary checks of the result in scalar-tensor sector. The Gauss-Bonnet corrections are also verified in the test-body limit for a companion with vanishing scalar sensitivity.

\subsubsection*{General-relativity limit}
\label{sec:gr-limit}

We first compare with general relativity by setting $s_1=s_2=0$, so that only the electric and magnetic response values contribute through 3PM; the angle must then reduce to the general-relativistic result~\cite{Kalin:2020lmz}, whose quadrupolar part agrees with~\cite{Cheung:2020sdj}. 
We match the tidal actions to relate the response coefficients to the couplings used in~\cite{Kalin:2020lmz}:
\begin{align}
 c^{(a)}_{E^2}=\frac{\rho_{a,0}^{(2)}}{4},\qquad
 c^{(a)}_{B^2}=\frac{2}{3}\,\xi_{a,0}^{(2)},\qquad
 c^{(a)}_{\tilde E^2}=-\frac{\rho_{a,0}^{(3)}}{12},\qquad
 c^{(a)}_{\tilde B^2}=-\frac{\xi_{a,0}^{(3)}}{4}.
 \label{eq:KLP-dictionary}
\end{align}
With this dictionary, the 2PM and 3PM terms agree exactly with that result, including those proportional to $\nu$ and those containing the inverse hyperbolic sine. The comparison thereby also tests the velocity dependence of the $m_1/m_2$ terms in $\mathcal B_\rho^{(\ell)}$ and $\mathcal B_\xi^{(\ell)}$, which is not captured by the test-body calculation.

\subsubsection*{Test-body limit}
\label{sec:probe}

We next consider $m_2\gg m_1$, where body~1 moves as a tidally deformed probe in the static scalar-tensor background generated by body~2, with the Einstein-frame metric and scalar field
\begin{align}
& \rmd s^2 =A(r)\rmd t^2-B(r)\rmd r^2-C(r)\rmd\Omega^2,
\nonumber\\[0.35 em]
& \varphi(r) =\varphi_0+\frac{s_2}{2\sqrt{1+s_2^2}}\log N,
 \label{eq:probe-background}
\end{align}
where
\begin{align}
 A(r)&=N^{1/\sqrt{1+s_2^2}},\qquad
 B(r)=N^{-1/\sqrt{1+s_2^2}},\nonumber\\
 C(r)&=r^2 N^{1-1/\sqrt{1+s_2^2}},\qquad
 N=1-\frac{2Gm_2\sqrt{1+s_2^2}}{r}.
 \label{eq:probe-metric-functions}
\end{align}
Setting $s_2=0$ recovers the Schwarzschild solution. The probe's mass and response coefficients are evaluated at the local scalar field $\varphi(r)$, and to linear order in the responses its on-shell condition reads
\begin{align}
 g^{\mu\nu}P_{1\mu}P_{1\nu}-M_1^2(\varphi)
 +2M_1(\varphi)\, \mathcal{O}_1[g,\varphi,P_1]&=0,
 \label{eq:probe-shell}
\end{align}
where $P_{1\mu}$ is the canonical momentum of the probe and $\mathcal{O}_1$ collects the tidal terms in its proper-time Lagrangian from~\eqref{eq:Sscalar}--\eqref{eq:Smixed}. Within $\mathcal{O}_1$, we replace $\dot{x}_1^\mu$ by $-g^{\mu\nu}P_{1\nu}/M_1(\varphi)$, retaining the background metric and scalar-dependent mass. Tidal corrections to this relation would enter $\mathcal{O}_1$ only at quadratic order in the response coefficients and are therefore omitted.

For equatorial motion, we set $\theta=\pi/2$ and $P_{1\theta}=0$; the momentum conjugate to the azimuthal angle is then the conserved orbital angular momentum $J=b_ep_\infty$. We solve~\eqref{eq:probe-shell} for the outgoing radial momentum $p_r\equiv P_{1r}$ perturbatively in $Gm_2/r$. The scattering angle follows from the radial action, differentiated with respect to $J$ at fixed probe energy~\cite{Damour:2016gwp,Almeida:2026abr}:
\begin{align}
	\chi&=-2\int_{r_t}^{\infty}\rmd r\,
	\frac{\partial p_r}{\partial J}-\pi,
\end{align}
where $r_t$ is the turning point of the tidally perturbed orbit. Expanding the integrand in $G$ and integrating term by term from the unperturbed turning point produce endpoint divergences; their Hadamard finite parts account for the shift of $r_t$, as in the post-Newtonian calculation of Appendix~\ref{app:pn-details}.

The resulting angle agrees with the $m_1/m_2\to0$ limit of~\eqref{eq:tidal-angle-result} for all retained responses of body~1, reproducing $\mathcal A_F^{(\ell)}$, $\mathcal C_F^{(\ell)}$, and the mass-ratio-independent part of $\mathcal B_F^{(\ell)}$. At $s_2=0$ and linear order in $\alpha$, the background acquires the scalar field $\varphi=\varphi_0-\alpha f_1G^2m_2^2/r^4+\mathcal O(G^3)$, sourced by the Gauss-Bonnet invariant of the Schwarzschild metric, and only the mixed responses couple to it at this order. In this background, the test-body calculation reproduces the Gauss-Bonnet terms in the mass-ratio-independent parts of $\mathcal B_\zeta^{(2)}$ and $\mathcal B_\zeta^{(3)}$ at $s_2=0$.

\subsubsection*{Post-Newtonian limit}
\label{sec:pn}

Finally, we test the mass and sensitivity dependence of the result against the post-Newtonian tidal dynamics~\cite{Bernard:2023eul}. Expanding~\eqref{eq:tidal-angle-result} in the asymptotic relative speed yields the 2PM contribution through absolute 5PN order,
\begin{align}
 \left.\Delta\chi\right|_{2\mathrm{PM}}
 &=\frac{\vartheta}{|\vartheta|}\nu
 \left(\frac{Gm}{b_e}\right)^4\Biggl\{
 \frac{m_1}{m}
 \frac{3\pi s_2^2\bar{\lambda}_{1,0}^{(1)}}{16}
 \left(\frac{4}{v^2}+2\nu-3-\frac{\nu^2}{2}v^2\right)
 \nonumber\\
 &\qquad+\left(\frac{Gm}{b_e}\right)^2\left(\frac{m_1}{m}\right)^3\frac{45\pi}{16v^2}
 \left(\bar{\rho}_{1,0}^{(2)}-s_2^2\bar{\lambda}_{1,0}^{(2)}
 -2s_2\bar{\zeta}_{1,0}^{(2)}\right)+(1\leftrightarrow2)\Biggr\}.
 \label{eq:PN-full-2PM}
\end{align}

The scalar dipole contributes from 3PN through 5PN, whereas the three quadrupolar responses enter at 5PN only through the combination $\bar{\rho}_{1,0}^{(2)}-s_2^2\bar{\lambda}_{1,0}^{(2)}-2s_2\bar{\zeta}_{1,0}^{(2)}$ and its counterpart for body~2. Nevertheless, the full 2PM coefficients of these three responses have distinct velocity dependences.

At 3PM, the expansion gives
\begin{align}
 \left.\Delta\chi\right|_{3\mathrm{PM}}
 &=\frac{\vartheta}{|\vartheta|}\nu^2
 \left(\frac{Gm}{b_e}\right)^5\Biggl\{
 \bar{\lambda}_{1,0}^{(1)}
 \Biggl[\frac{8m}{m_2}\frac{s_2^2(1+s_1s_2)}{v^4}
 \nonumber\\
 &\quad-\frac{4s_2}{3v^2}\Bigl\{
 \frac{m_1}{m_2}\bigl[(15-3\nu)s_1s_2^2
 +4s_1s'_2-(5+3\nu)s_2\bigr]
 \nonumber\\
 &\hspace{41mm}+(12-3\nu)s_1s_2^2-(2+3\nu)s_2\Bigr\}
 \nonumber\\
 &\quad+\frac{s_2}{15}\Bigl\{
 \frac{m_1}{m_2}\bigl[(105-105\nu-15\nu^2)s_1s_2^2
 +40(1-\nu)s_1s'_2
 \nonumber\\
 &\hspace{47mm}-(111-95\nu+15\nu^2)s_2\bigr]
 \nonumber\\
 &\hspace{28mm}+(128-75\nu-15\nu^2)s_1s_2^2
 -(96-65\nu+15\nu^2)s_2\Bigr\}\Biggr]
 \nonumber\\
 &\quad+\frac{4s_2^3\bar{\lambda}_{1,1}^{(1)}}{15}
 \left(8-5\nu-\frac{10}{v^2}\right)
 \nonumber\\
 &\quad+\left(\frac{Gm}{b_e}\right)^2\left(\frac{m_1}{m}\right)^2
 \frac{48m(1+s_1s_2)}{m_2v^4}
 \left(\bar{\rho}_{1,0}^{(2)}-s_2^2\bar{\lambda}_{1,0}^{(2)}
 -2s_2\bar{\zeta}_{1,0}^{(2)}\right)
 +(1\leftrightarrow2)\Biggr\}.
 \label{eq:PN-full-3PM}
\end{align}
In the scalar-dipole term proportional to the response value, the powers $v^{-4},v^{-2},v^0$ correspond to 3PN, 4PN, and 5PN, respectively. The response derivative contributes at 4PN and 5PN.

We also independently derive the angle from the post-Newtonian conserved energy and angular momentum~\cite{Bernard:2023eul,Bernard:2018ivi}, as described in Appendix~\ref{app:pn-details}. All overlapping 2PM and 3PM contributions agree through absolute 5PN order, including the vanishing 2PM terms proportional to $\bar{\lambda}_{a,1}^{(1)}$ at 4PN and 5PN. The magnetic and octupolar responses, as well as the first scalar derivatives of the quadrupolar responses, enter only beyond 5PN.

\begin{table}[t]
\centering
\begin{tabular}{@{}lll@{}}
\toprule
Response coefficients & 2PM & 3PM\\
\midrule
$\bar{\lambda}_{1,0}^{(1)}$ & 3PN, 4PN, 5PN & 3PN, 4PN, 5PN\\
$\bar{\lambda}_{1,1}^{(1)}$ & 4PN, 5PN (both vanish) & 4PN, 5PN\\
$\bar{\rho}_{1,0}^{(2)}$, $\bar{\lambda}_{1,0}^{(2)}$, $\bar{\zeta}_{1,0}^{(2)}$ & 5PN & 5PN\\
\bottomrule
\end{tabular}
\caption{Post-Newtonian comparisons for body~1. Each entry lists the PN orders compared for the corresponding coefficient in the 2PM or 3PM expansion; the last row includes three coefficients per PM order. The 16 comparisons per body include 14 nonzero coefficients.}
\label{tab:pn-comparisons}
\end{table}

\section{Conclusions}
\label{sec:conclusions}

We have computed the conservative tidal contributions to the impulse and scattering angle through 3PM for nonspinning bodies in massless scalar-tensor theories with a scalar-Gauss-Bonnet coupling. Our results retain the full velocity dependence at linear order in the response coefficients, cover the scalar dipole and quadrupole as well as the electric, magnetic, and mixed quadrupoles and octupoles, and reproduce the general-relativistic, test-body, and post-Newtonian limits. They extend the general-relativistic tidal calculation~\cite{Kalin:2020lmz} and complement the 3PM results without tidal effects~\cite{Almeida:2026abr}.

At 2PM, only the response values at $\varphi_0$ appear; at 3PM, their first scalar derivatives also enter, together with explicit Gauss-Bonnet terms linear in $\alpha f_1$. Organized by the mass ratio, each coefficient $\mathcal B_F^{(\ell)}$ consists of a term independent of $m_1/m_2$, which is rational in $\gamma$, and a term linear in this ratio, which contains all logarithms.

Through the boundary-to-bound correspondence~\cite{Kalin:2019rwq,Kalin:2019inp}, the scattering data can also be used to derive bound-orbit observables for adiabatic tides, such as the circular-orbit binding energy. The same results can inform the tidal sector of effective-one-body models of scalar-tensor binaries~\cite{Damour:2009wj,Julie:2017ucp,Jain:2022nxs}. Finally, extending the test-body calculation to $s_2\neq0$ at $\mathcal O(\alpha)$ would check the remaining Gauss-Bonnet terms in the mass-ratio-independent part of $\mathcal B_F^{(\ell)}$.

\acknowledgments
We thank Gabriel Luz Almeida for fruitful collaboration on related work, numerous helpful discussions, and valuable comments on the manuscript.
This work was supported in part by the Start-up Research Fund of Southeast University (No.\,RF1028624160).

\appendix
\section{Conventions and dictionaries}
\label{app:conventions}

Comparing tidal results requires consistent conventions for the frame, metric signature, and normalization of both the scalar field and the responses. The dictionaries below use the dimensionful response coefficients of the worldline action, which are related to the dimensionless responses by~\eqref{eq:dimensionless-responses}. Symbols taken from other references retain their original definitions.

\subsection{Jordan-frame dictionary}
\label{app:laura-notation}

Converting the post-Newtonian tidal results in~\cite{Bernard:2023eul} from the Jordan frame to our conventions mixes the quadrupolar response coefficients. We use the definitions in equations (7)--(18) and (57)--(60) and Table~III of that reference, which adopts the mostly-plus signature. As in the post-Newtonian comparison, we set $c=1$ and $\phi_{{\rm J}0}=1$, and work at $\alpha=0$. Appendix~\ref{app:canonical} discusses canonical normalization.

\subsubsection*{Field and response transformations}

The transformation of the response derivatives depends on the scalar normalization. We denote the Jordan scalar by $\phi_{\rm J}$ and the normalized field by
$\varphi_{\rm J}=\phi_{\rm J}/\phi_{{\rm J}0}$.
On the Jordan-frame side, the Taylor coefficients are defined as derivatives with respect to the Jordan scalar:
\begin{align}
 F_{a,{\rm J}}(\phi_{\rm J})&=
 \sum_{k\geq0}\frac{F_{a,{\rm J}}^{(k)}}{k!}
 \phi_{{\rm J}0}^{k}(\varphi_{\rm J}-1)^{k},\qquad
 F_{a,{\rm J}}^{(k)}=
 \left.\frac{\rmd^{k}F_{a,{\rm J}}}{\rmd\phi_{\rm J}^{k}}
 \right|_{\phi_{{\rm J}0}}.
 \label{eq:app-Jordan-Taylor}
\end{align}
Derivatives with respect to $\varphi_{\rm J}$ therefore carry a factor $\phi_{{\rm J}0}^{k}$ before the background normalization is imposed. In the Jordan-frame notation, $\lambda,\mu,\nu,c$ denote the scalar dipole, scalar quadrupole, mixed quadrupole, and electric quadrupole responses, respectively.

In our signature, the two metrics are related by the conformal factor $A_{\mathrm c}(\varphi)$, with $A_{\mathrm c0}=A_{\mathrm c}(\varphi_0)=1$:
\begin{align}
 g^{\rm J}_{\mu\nu}&=A_{\mathrm c}^2(\varphi)g_{\mu\nu},\qquad
 A_{\mathrm c}=\varphi_{\rm J}^{-1/2},\qquad
 \rmd\tau_{\rm J}=A_{\mathrm c}\,\rmd\tau.
 \label{eq:app-frame-map}
\end{align}
The logarithmic derivative of the conformal factor, fixed by matching the scalar kinetic terms, and the scalar Jacobian are
\begin{align}
&
 \alpha_{\mathrm c}(\varphi) =\frac{\rmd\ln A_{\mathrm c}}{\rmd\varphi},\qquad
 \alpha_{\mathrm c0}=\alpha_{\mathrm c}(\varphi_0),\qquad
 \beta_0=\alpha_{\mathrm c}'(\varphi_0),\qquad \beta'_0=\alpha_{\mathrm c}''(\varphi_0),
 \nonumber\\
&
 \frac{\rmd\ln\varphi_{\rm J}}{\rmd\varphi} =-2\alpha_{\mathrm c},\qquad
 \alpha_{\mathrm c}^2=\frac{1}{3+2\omega},\qquad
 \frac{\rmd}{\rmd\varphi_{\rm J}}
 =-\frac{A_{\mathrm c}^2}{2\alpha_{\mathrm c}}\frac{\rmd}{\rmd\varphi},
 \label{eq:app-scalar-Jacobian}
\end{align}
where $\omega$ is the Jordan-frame coupling function. The inverse scalar map requires $\alpha_{\mathrm c}\ne0$ locally and $\alpha_{\mathrm c0}\ne0$ at the background. We retain the sign of $\alpha_{\mathrm c0}$ and use the full Einstein-frame scalar $\varphi=\varphi_0+\kappa\phi$. The scalar-dependent masses transform according to
\begin{align}
 M_a(\varphi)&=A_{\mathrm c}(\varphi)
 m_{a,{\rm J}}\bigl(\phi_{\rm J}(\varphi)\bigr),\qquad
 \frac{\rmd}{\rmd\varphi}
 =-2\alpha_{\mathrm c}\frac{\rmd}{\rmd\ln\varphi_{\rm J}}
 \quad\hbox{on Jordan functions}.
 \label{eq:app-mass-map}
\end{align}
In particular, $s_a=\alpha_{\mathrm c0}(1-2s_{a,{\rm J}})$.
Further differentiation yields the sensitivity conversions listed in the summary below. Primes on Jordan-frame sensitivities denote derivatives with respect to $\ln\phi_{\rm J}$, whereas $s'_a,s''_a$ retain the definitions in \eqref{eq:mass-sensitivities}. With this background normalization, each body has the same asymptotic mass in both frames.

The dipole conversion combines the proper-time factor $A_{\mathrm c}$, the inverse-metric factor $A_{\mathrm c}^{-2}$, and two scalar Jacobians, $(-2\alpha_{\mathrm c}\varphi_{\rm J})^2$.
Changing from the $(-+++)$ signature to our $(+---)$ signature reverses the scalar-gradient square, so
\begin{align}
 \lambda_{a}^{(1)}(\varphi)&=-4\alpha_{\mathrm c}^2A_{\mathrm c}^{-1}\varphi_{\rm J}^2
 \lambda_{a,{\rm J}}(\phi_{\rm J})
 =-4\alpha_{\mathrm c}^2A_{\mathrm c}^{-5}\lambda_{a,{\rm J}}(\phi_{\rm J}).
 \label{eq:app-dipole-function}
\end{align}
The second equality uses $\varphi_{\rm J}^2=A_{\mathrm c}^{-4}$.
Differentiating the full transformed function yields the Taylor coefficients; for example,
\begin{align}
 \lambda_{a,1}^{(1)}&=
 \left(\frac{2\beta_0}{\alpha_{\mathrm c0}}-5\alpha_{\mathrm c0}\right)\lambda_{a,0}^{(1)}
 +8\alpha_{\mathrm c0}^3\lambda_{a,{\rm J}}^{(1)}.
 \label{eq:app-dipole-first-derivative}
\end{align}
A second differentiation yields the second-derivative conversion.

At the next-to-next-to-leading post-Newtonian order retained here, the reduced Riemann basis defined in formula~(10) in~\cite{Bernard:2023eul} yields the relations below. The label L marks quantities in the signature and conventions of that reference:
\begin{align}
&
 (\nabla^\perp_{\mu\nu}\varphi_{\rm J})_{{\rm J},{\rm L}}
=-2\alpha_{\mathrm c0}(\nabla^\perp_{\mu\nu}\varphi)_{\rm L}+\cdots,
\nonumber\\[0.5 em]
&
 G^{{\rm J},{\rm L}}_{\mu\nu} =G^{\rm L}_{\mu\nu}
 -\alpha_{\mathrm c0}(\nabla^\perp_{\mu\nu}\varphi)_{\rm L}+\cdots .
 \label{eq:app-quadrupole-moments}
\end{align}
In this basis, the linear conformal transformation generates the spatial scalar Hessian. The time-derivative trace is suppressed by $v^2$, while scalar-gradient products and the Einstein-frame Ricci tensor carry an additional field. These terms therefore enter beyond the leading quadrupolar contribution at 5PN.

Inserting these relations into the scalar, electric, and mixed quadratic operators and using $G^{\rm L}_{\mu\nu}=-G_{\mu\nu}$ yields
\begin{align}
 \rho_{a,0}^{(2)}&=c_{a,{\rm J}}^{(0)},\nonumber\\
 \zeta_{a,0}^{(2)}&=\alpha_{\mathrm c0}\left(2\nu_{a,{\rm J}}^{(0)}-c_{a,{\rm J}}^{(0)}\right),\nonumber\\
 \lambda_{a,0}^{(2)}&=\alpha_{\mathrm c0}^2\left(4\mu_{a,{\rm J}}^{(0)}
 +4\nu_{a,{\rm J}}^{(0)}-c_{a,{\rm J}}^{(0)}\right).
 \label{eq:app-quadrupole-forward}
\end{align}
The sign in $G^{\rm L}_{\mu\nu}=-G_{\mu\nu}$ follows from lowering the first curvature index with the opposite metric signature while keeping the Riemann convention fixed.

The three quadrupolar responses must therefore be converted together. This map holds at the retained post-Newtonian order, and we do not extend it to Weyl moments beyond this order. The post-Newtonian comparison requires only the responses included here, so magnetic and octupolar maps are unnecessary.
The second-derivative dipole relation in the summary below is an algebraic response-coefficient conversion. The $k=2$ coefficient first contributes to the scattering angle at 4PM and therefore does not enter the post-Newtonian comparison.

\subsubsection*{Conversion summary}

In the Jordan-frame relations below, the quantities on the left-hand sides follow the definitions of Bernard, Dones, and Mougiakakos~\cite{Bernard:2023eul}, whereas those on the right-hand sides are expressed in our conventions. For the responses, parenthesized superscripts count scalar derivatives on the Jordan-frame side and multipole order on our side, while our second subscript counts scalar derivatives.
We set
$\phi_{{\rm J}0}=A_{\mathrm c0}=1$, $\alpha=0$, and $\alpha_{\mathrm c0}\ne0$.
The masses satisfy $m_{a,{\rm J}}(1)=m_a$.

\noindent\textit{Sensitivities:}
\begin{align*}
s_{a,{\rm J}}&=\frac12\left(1-\frac{s_a}{\alpha_{\mathrm c0}}\right) 
\\[0.3em]
s'_{a,{\rm J}}&=\frac{s'_a-\beta_0s_a/\alpha_{\mathrm c0}}{4\alpha_{\mathrm c0}^2} 
\\[0.3em]
s''_{a,{\rm J}}&=\frac{1}{8\alpha_{\mathrm c0}^3}\left[-s''_a+\frac{3\beta_0}{\alpha_{\mathrm c0}}s'_a
 +\left(\frac{\beta'_0}{\alpha_{\mathrm c0}}-\frac{3\beta_0^2}{\alpha_{\mathrm c0}^2}\right)s_a\right]
\end{align*}
\noindent\textit{Scalar dipole response:}
\begin{align*}
\lambda_{a,{\rm J}}^{(0)}&=-\frac{\lambda_{a,0}^{(1)}}{4\alpha_{\mathrm c0}^2} 
\\[0.3em]
\lambda_{a,{\rm J}}^{(1)}&=\frac{1}{8\alpha_{\mathrm c0}^3}
\left[\lambda_{a,1}^{(1)}+\left(5\alpha_{\mathrm c0}-\frac{2\beta_0}{\alpha_{\mathrm c0}}\right)\lambda_{a,0}^{(1)}\right] 
\\[0.3em]
\lambda_{a,{\rm J}}^{(2)}&=-\frac{1}{16\alpha_{\mathrm c0}^4}\Biggl[\lambda_{a,2}^{(1)}
 +\left(12\alpha_{\mathrm c0}-\frac{5\beta_0}{\alpha_{\mathrm c0}}\right)\lambda_{a,1}^{(1)}
 +\left(35\alpha_{\mathrm c0}^2-24\beta_0+\frac{8\beta_0^2}{\alpha_{\mathrm c0}^2}
 -\frac{2\beta'_0}{\alpha_{\mathrm c0}}\right)\lambda_{a,0}^{(1)}\Biggr]
\end{align*}
\noindent\textit{Quadrupolar responses:}
\begin{align*}
c_{a,{\rm J}}^{(0)}&=\rho_{a,0}^{(2)} \\[0.3em]
\nu_{a,{\rm J}}^{(0)}&=\frac{\zeta_{a,0}^{(2)}+\alpha_{\mathrm c0}\rho_{a,0}^{(2)}}{2\alpha_{\mathrm c0}} \\[0.3em]
\mu_{a,{\rm J}}^{(0)}&=\frac{\lambda_{a,0}^{(2)}-2\alpha_{\mathrm c0}\zeta_{a,0}^{(2)}-\alpha_{\mathrm c0}^2\rho_{a,0}^{(2)}}{4\alpha_{\mathrm c0}^2}
\end{align*}

\subsection{Canonical normalization}
\label{app:canonical}
For the scalar normalization~\eqref{eq:scalar-expansion} used in our results, the flat-space kinetic term reads
\begin{align}
 \mathcal L_{\phi,\rm kin}&=4\,\partial_\mu\phi\partial^\mu\phi
 =\frac12\partial_\mu\psi\partial^\mu\psi,\qquad
 \psi=2\sqrt2\,\phi.
 \label{eq:app-canonical-field}
\end{align}
The canonical field $\psi$ should not be confused with the post-Newtonian perturbation $\psi_{\rm L}=\varphi_{\rm J}-1$ used in the Jordan-frame calculation.
For each multipole $\ell$, the expansion in the canonical field defines
\begin{align}
 F_a^{(\ell),\mathrm{can}}(\psi)=\sum_{k}F_{a,k}^{(\ell),\mathrm{can}}(\kappa\psi)^{k}/k!.
 \label{eq:canonical-coefficients}
\end{align}
The corresponding coefficients are related by
\begin{align*}
&
\lambda_{a,k}^{(\ell),\mathrm{can}} = \frac{\lambda_{a,k}^{(\ell)}}{(2\sqrt2)^{k+2}},
\qquad  \zeta_{a,k}^{(\ell),\mathrm{can}}=\frac{\zeta_{a,k}^{(\ell)}}{(2\sqrt2)^{k+1}},
\\[0.35em]
&
\rho_{a,k}^{(\ell),\mathrm{can}} =\frac{\rho_{a,k}^{(\ell)}}{(2\sqrt2)^{k}}, 
\qquad\quad \xi_{a,k}^{(\ell),\mathrm{can}} =\frac{\xi_{a,k}^{(\ell)}}{(2\sqrt2)^{k}},
\qquad\quad f_{k}^{\rm can}=\frac{f_{k}}{(2\sqrt2)^{k}}.
\end{align*}
At Taylor order $k$, the response function contributes $k$ scalar fields, while the operator supplies two more for $\lambda^{(\ell)}$, one for $\zeta^{(\ell)}$, and none for $\rho^{(\ell)},\xi^{(\ell)}$. This counting fixes the powers in the conversion factors independently of $\ell$. The rescaling preserves the factorial factor $1/k!$ in the Taylor expansion and introduces no sign change. In a vertex, differentiating the Taylor factor supplies a factor of $k!$; the explicit scalar fields in the operator must also be symmetrized over the external legs.

\section{Coefficients}
\label{app:tidal-coefficients}

The coefficient functions below determine the tidal contribution to the impulse from a response on body~1; exchanging the body labels yields the contribution from body~2. We use the variable $x$ defined in~\eqref{eq:tidal-x-variable} and the polynomials $h_i(\gamma)$ listed at the end of this appendix.

\subsection{2PM coefficients}
\label{app:2pm-coefficients}

The eight 2PM coefficients are
\begin{align}
 \mathcal A_{\lambda}^{(1)}&=\frac{3\pi {s_2}^{2}\bigl(\gamma^{2}+3\bigr)}{16(\gamma^2-1)},\qquad \mathcal A_{\lambda}^{(2)}=-\frac{45\pi {s_2}^{2}h_{12}}{512(\gamma^2-1)},\nonumber\\
 \mathcal A_{\rho}^{(2)}&=\frac{45\pi h_{21}}{256(\gamma^2-1)},\qquad \mathcal A_{\rho}^{(3)}=-\frac{175\pi h_{50}}{2048(\gamma^2-1)},\nonumber\\
 \mathcal A_{\xi}^{(2)}&=\frac{75\pi \bigl(7\gamma^{2}+1\bigr)}{32},\qquad \mathcal A_{\xi}^{(3)}=-\frac{3675\pi h_{4}}{2048},\nonumber\\
 \mathcal A_{\zeta}^{(2)}&=-\frac{45\pi s_2\bigl(3\gamma^{2}-1\bigr)}{16(\gamma^2-1)},\qquad \mathcal A_{\zeta}^{(3)}=\frac{525\pi s_2h_{5}}{256(\gamma^2-1)}.
 \label{eq:tidal-2PM-coefficients}
\end{align}

\subsection{3PM coefficients}
\label{app:3pm-coefficients}

The coefficients multiplying the first scalar derivatives of the responses are
\begin{align}
 \mathcal C_{\lambda}^{(1)}&=-\frac{8{s_2}^{3}\bigl(\gamma^{2}+4\bigr)}{15(\gamma^2-1)},\qquad \mathcal C_{\lambda}^{(2)}=\frac{16{s_2}^{3}h_{6}}{35(\gamma^2-1)},\nonumber\\
 \mathcal C_{\rho}^{(2)}&=-\frac{48s_2h_{13}}{35(\gamma^2-1)},\qquad \mathcal C_{\rho}^{(3)}=\frac{128s_2h_{63}}{9009(\gamma^2-1)},\nonumber\\
 \mathcal C_{\xi}^{(2)}&=-\frac{256s_2\bigl(8\gamma^{2}+1\bigr)}{35},\qquad \mathcal C_{\xi}^{(3)}=\frac{2048s_2h_{23}}{3003},\nonumber\\
 \mathcal C_{\zeta}^{(2)}&=\frac{32{s_2}^{2}\bigl(32\gamma^{2}-11\bigr)}{35(\gamma^2-1)},\qquad \mathcal C_{\zeta}^{(3)}=-\frac{256{s_2}^{2}h_{37}}{1155(\gamma^2-1)}.
 \label{eq:tidal-derivative-coefficients}
\end{align}

The coefficients multiplying the responses at the asymptotic scalar value are grouped by their mass-ratio dependence. The first bracket in each expression is independent of $m_1/m_2$; the second is proportional to this ratio.
\begin{align}
\mathcal B_\lambda^{(1)}
={}&\frac{8s_2}{3(\gamma^2-1)^2}\Biggl\{
\Biggl[\frac{s_2}{5}(s_1s_2h_2+h_{14})
+\frac{48\alpha f_1}{35b_e^2}(\gamma^2-1)(\gamma^2+6)\Biggr]
\nonumber\\[0.5em]
&\quad+\frac{m_1}{m_2}\Biggl[
\gamma\Bigl(s_1\bigl[s_2^2(\gamma^2+32)+15s'_2+30(\gamma^2+2)\bigr]+s_2h_1\Bigr)
\nonumber\\
&\qquad+\frac{3\log x}{\sqrt{\gamma^2-1}}
\Bigl(s_1\bigl[(2s_2^2+s'_2)(4\gamma^2+1)+2h_7\bigr]+2s_2(\gamma^2+1)\Bigr)
\nonumber\\
&\qquad-\frac{12\alpha f_1}{b_e^2(\gamma^2-1)}
\left(\gamma h_{24}+\frac{3h_{25}\log x}{\sqrt{\gamma^2-1}}\right)
\Biggr]\Biggr\}.
 \label{eq:tidal-B-lambda110}
\end{align}
\begin{align}
\mathcal B_\lambda^{(2)}
={}&-\frac{2s_2}{5(\gamma^2-1)^3}\Biggl\{
\Biggl[\frac{8s_2}{7}(\gamma^2-1)(s_1s_2h_{44}+h_{45})
+\frac{512\alpha f_1}{77b_e^2}(\gamma^2-1)^2h_{15}\Biggr]
\nonumber\\[0.5em]
&\quad+\frac{m_1}{m_2}\Biggl[
\gamma\Bigl(s_1\bigl[2s_2^2h_{53}+15s'_2(266\gamma^2+235)+20h_{41}\bigr]+4s_2h_{71}\Bigr)
\nonumber\\
&\qquad+\frac{15\log x}{\sqrt{\gamma^2-1}}
\Bigl(s_1\bigl[2s_2^2h_{31}+s'_2h_{32}+4h_{51}\bigr]+4s_2h_{26}\Bigr)
\nonumber\\
&\qquad-\frac{240\alpha f_1}{b_e^2(\gamma^2-1)}
\left(\gamma h_{69}+\frac{3h_{77}\log x}{\sqrt{\gamma^2-1}}\right)
\Biggr]\Biggr\}.
 \label{eq:tidal-B-lambda210}
\end{align}
\begin{align}
\mathcal B_\rho^{(2)}
={}&\frac{12}{5(\gamma^2-1)^3}\Biggl\{
\Biggl[\frac{4(\gamma^2-1)}{21}
\bigl[3(s_1s_2h_{46}+h_{47})-2s_2^2(\gamma^2-1)h_{13}\bigr]
\nonumber\\
&\hspace{20mm}
+\frac{256\alpha f_1s_2}{77b_e^2}(\gamma^2-1)^2h_{22}\Biggr]
\nonumber\\
&+\frac{m_1}{m_2}\Biggl[
2\gamma(s_1s_2h_{48}+h_{74})
+\frac{30\log x}{\sqrt{\gamma^2-1}}
\bigl[h_{33}+5s_1s_2(4\gamma^2+1)\bigr]
\nonumber\\
&\hspace{15mm}
+\frac{80\alpha f_1}{b_e^2(\gamma^2-1)}
\Bigl(\gamma\bigl[s_1h_{78}-6s_2(37\gamma^2+38)\bigr]
\nonumber\\
&\hspace{48mm}
-\frac{3\log x}{\sqrt{\gamma^2-1}}(s_1h_{64}+6s_2h_8)
\Bigr)\Biggr]\Biggr\}.
 \label{eq:tidal-B-c210}
\end{align}
\begin{align}
\mathcal B_\rho^{(3)}
={}&-\frac{32}{21(\gamma^2-1)^4}\Biggl\{
\Biggl[\frac{4(\gamma^2-1)^2}{2145}
\bigl[5(s_1s_2h_{85}+h_{82})-4s_2^2(\gamma^2-1)h_{63}\bigr]
\nonumber\\
&\hspace{20mm}
+\frac{512\alpha f_1s_2}{143b_e^2}(\gamma^2-1)^3h_{54}\Biggr]
\nonumber\\
&+\frac{m_1}{m_2}\Biggl[
\gamma(s_1s_2h_{90}+h_{96})
+\frac{105\log x}{\sqrt{\gamma^2-1}}(h_{66}+s_1s_2h_{40})
\nonumber\\
&\hspace{15mm}
+\frac{70\alpha f_1}{b_e^2(\gamma^2-1)}
\Bigl(\gamma(s_1h_{94}-30s_2h_{60})
\nonumber\\
&\hspace{48mm}
-\frac{15\log x}{\sqrt{\gamma^2-1}}(s_1h_{86}+6s_2h_{59})
\Bigr)\Biggr]\Biggr\}.
 \label{eq:tidal-B-c310}
\end{align}
\begin{align}
\mathcal B_\xi^{(2)}
={}&\frac{32}{5(\gamma^2-1)^3}\Biggl\{
\Biggl[\frac{8(\gamma^2-1)^2}{21}
\bigl[3(3s_1s_2h_9+h_{16})-2s_2^2(\gamma^2-1)(8\gamma^2+1)\bigr]
\nonumber\\
&\hspace{20mm}
+\frac{2048\alpha f_1s_2}{77b_e^2}(\gamma^2-1)^3(10\gamma^2+1)\Biggr]
\nonumber\\
&+\frac{m_1}{m_2}\Biggl[
2\gamma(s_1s_2h_{49}+2h_{73})
+\frac{30\log x}{\sqrt{\gamma^2-1}}
\bigl[2h_{34}+s_1s_2(20\gamma^2+3)\bigr]
\nonumber\\
&\hspace{15mm}
+\frac{640\alpha f_1s_1}{b_e^2(\gamma^2-1)}
\Bigl(\gamma h_{72}-\frac{3h_{56}\log x}{\sqrt{\gamma^2-1}}\Bigr)
\Biggr]\Biggr\}.
 \label{eq:tidal-B-d210}
\end{align}
\begin{align}
\mathcal B_\xi^{(3)}
={}&-\frac{32}{7(\gamma^2-1)^4}\Biggl\{
\Biggl[\frac{64(\gamma^2-1)^3}{2145}
\bigl[5(s_1s_2h_{61}+h_{57})-4s_2^2(\gamma^2-1)h_{23}\bigr]
\nonumber\\
&\hspace{20mm}
+\frac{20480\alpha f_1s_2}{143b_e^2}(\gamma^2-1)^4h_3\Biggr]
\nonumber\\
&+\frac{m_1}{m_2}\Biggl[
\gamma(s_1s_2h_{91}+h_{97})
+\frac{105\log x}{\sqrt{\gamma^2-1}}(h_{67}+s_1s_2h_{36})
\nonumber\\
&\hspace{15mm}
+\frac{70\alpha f_1s_1}{b_e^2(\gamma^2-1)}
\Bigl(\gamma h_{95}-\frac{15h_{87}\log x}{\sqrt{\gamma^2-1}}\Bigr)
\Biggr]\Biggr\}.
 \label{eq:tidal-B-d310}
\end{align}
\begin{align}
\mathcal B_\zeta^{(2)}
={}&\frac{8}{(\gamma^2-1)^3}\Biggl\{
\Biggl[\frac{4s_2}{105}(\gamma^2-1)
\Bigl[s_2^2(\gamma^2-1)(32\gamma^2-11)-3(s_1s_2h_{29}+3h_{19})\Bigr]
\nonumber\\
&\qquad-\frac{128\alpha f_1}{35b_e^2}(\gamma^2-1)^2
\Bigl[s_2^2(14\gamma^2-5)+3(3\gamma^2-1)\Bigr]\Biggr]
\nonumber\\[0.5em]
&\quad+\frac{m_1}{m_2}\Biggl[
\gamma\Bigl(s_1\bigl[4s_2^2h_{27}+6s'_2h_{17}-3h_{28}\bigr]-8s_2h_{42}\Bigr)
\nonumber\\
&\qquad-\frac{3\log x}{\sqrt{\gamma^2-1}}
\Bigl(s_1\bigl[4s_2^2h_{18}+6s'_2(8\gamma^2+1)+15h_{10}\bigr]+12s_2h_{11}\Bigr)
\nonumber\\
&\qquad+\frac{24\alpha f_1}{b_e^2(\gamma^2-1)}\Biggl(
\gamma\Bigl[s_1s_2h_{68}+4(s_2^2h_{35}+h_{39})\Bigr]
\nonumber\\
&\hspace{49mm}
+\frac{3\log x}{\sqrt{\gamma^2-1}}
\Bigl[s_1s_2h_{76}+4(s_2^2h_{43}+h_{52})\Bigr]
\Biggr)\Biggr]\Biggr\}.
 \label{eq:tidal-B-v210}
\end{align}
\begin{align}
\mathcal B_\zeta^{(3)}
={}&\frac{16}{5(\gamma^2-1)^4}\Biggl\{
\Biggl[\frac{16s_2}{45045}(\gamma^2-1)^2
\Bigl[195s_1s_2h_{65}-78s_2^2(\gamma^2-1)h_{37}+5h_{83}\Bigr]
\nonumber\\
&\qquad+\frac{1024\alpha f_1}{3003b_e^2}(\gamma^2-1)^3
(3s_2^2h_{38}+65h_{20})\Biggr]
\nonumber\\[0.5em]
&\quad+\frac{m_1}{m_2}\Biggl[
\gamma\Bigl(s_1\bigl[2s_2^2h_{80}-15s'_2h_{58}-40h_{75}\bigr]+2s_2h_{92}\Bigr)
\nonumber\\
&\qquad+\frac{15\log x}{\sqrt{\gamma^2-1}}
\Bigl(s_1\bigl[2s_2^2h_{62}+15s'_2h_{30}+40h_{55}\bigr]+2s_2h_{79}\Bigr)
\nonumber\\
&\qquad-\frac{20\alpha f_1}{b_e^2(\gamma^2-1)}\Biggl(
\gamma\Bigl[s_1s_2h_{89}+5(s_2^2h_{70}+h_{88})\Bigr]
\nonumber\\
&\hspace{49mm}
+\frac{15\log x}{\sqrt{\gamma^2-1}}
(s_1s_2h_{93}+s_2^2h_{81}+5h_{84})
\Biggr)\Biggr]\Biggr\}.
 \label{eq:tidal-B-v310}
\end{align}
\begingroup
\allowdisplaybreaks[4]

\subsection{Polynomials}
\label{app:polynomials}

The polynomials $h_i(\gamma)$ appearing in the coefficients above are grouped by length, with short expressions first.
\begin{flalign*}
h_{1}&=2\gamma^{4}+\gamma^{2}+12 & h_{2}&=\gamma^{4}-2\gamma^{2}+16 \\
h_{3}&=\gamma^{4}+24\gamma^{2}+1 & h_{4}&=3\gamma^{4}+58\gamma^{2}+3 \\
h_{5}&=5\gamma^{4}+18\gamma^{2}-7 & h_{6}&=2\gamma^{4}+6\gamma^{2}+13 \\
h_{7}&=2\gamma^{4}+11\gamma^{2}+2 & h_{8}&=6\gamma^{4}+17\gamma^{2}+2 \\
h_{9}&=8\gamma^{4}+11\gamma^{2}+2 & h_{10}&=8\gamma^{4}+12\gamma^{2}+1 \\
h_{11}&=8\gamma^{4}+23\gamma^{2}+3 & h_{12}&=3\gamma^{4}+10\gamma^{2}+19 \\
h_{13}&=16\gamma^{4}-14\gamma^{2}+5 & h_{14}&=8\gamma^{4}+19\gamma^{2}-12 \\
h_{15}&=3\gamma^{4}+16\gamma^{2}+47 & h_{16}&=80\gamma^{4}-16\gamma^{2}-1 \\
h_{17}&=2\gamma^{4}-10\gamma^{2}-19 & h_{18}&=12\gamma^{4}+61\gamma^{2}+8 \\
h_{19}&=88\gamma^{4}-57\gamma^{2}+4 & h_{20}&=5\gamma^{4}+27\gamma^{2}-10 \\
h_{21}&=35\gamma^{4}-30\gamma^{2}+11 & h_{22}&=80\gamma^{4}-72\gamma^{2}+25 \\
h_{23}&=25\gamma^{4}+522\gamma^{2}+25 & h_{24}&=8\gamma^{4}+410\gamma^{2}+311 \\
h_{25}&=76\gamma^{4}+152\gamma^{2}+15 & h_{26}&=32\gamma^{4}+121\gamma^{2}+15 \\
h_{27}&=2\gamma^{4}-100\gamma^{2}-145 & h_{28}&=8\gamma^{4}+194\gamma^{2}+113 \\
h_{29}&=32\gamma^{4}+139\gamma^{2}-66 & h_{30}&=144\gamma^{4}+124\gamma^{2}+7 \\
h_{31}&=136\gamma^{4}+336\gamma^{2}+41 & h_{32}&=136\gamma^{4}+328\gamma^{2}+37 \\
h_{33}&=440\gamma^{4}+474\gamma^{2}+33 & h_{34}&=220\gamma^{4}+237\gamma^{2}+16 \\
h_{35}&=20\gamma^{4}+223\gamma^{2}+132 & h_{36}&=814\gamma^{4}+723\gamma^{2}+40 \\
h_{37}&=150\gamma^{4}+569\gamma^{2}-224 & h_{38}&=206\gamma^{4}+849\gamma^{2}-340 \\
h_{39}&=674\gamma^{4}+2215\gamma^{2}+636 & h_{40}&=1126\gamma^{4}+1163\gamma^{2}+74 \\
h_{41}&=508\gamma^{4}+2584\gamma^{2}+1003 & h_{42}&=4\gamma^{6}-6\gamma^{4}+75\gamma^{2}+80 \\
h_{43}&=2\gamma^{6}+48\gamma^{4}+69\gamma^{2}+6 & h_{44}&=6\gamma^{6}+10\gamma^{4}+11\gamma^{2}+78 \\
h_{45}&=16\gamma^{6}+50\gamma^{4}+62\gamma^{2}-23 & h_{46}&=48\gamma^{6}+18\gamma^{4}-61\gamma^{2}+30 \\
h_{47}&=160\gamma^{6}-192\gamma^{4}+72\gamma^{2}-5 & h_{48}&=16\gamma^{6}-28\gamma^{4}+68\gamma^{2}+319 \\
h_{49}&=16\gamma^{6}-28\gamma^{4}+78\gamma^{2}+279 & h_{50}&=21\gamma^{6}+385\gamma^{4}-305\gamma^{2}+91 \\
h_{51}&=64\gamma^{6}+656\gamma^{4}+606\gamma^{2}+39 & h_{52}&=94\gamma^{6}+630\gamma^{4}+429\gamma^{2}+22 \\
h_{53}&=8\gamma^{6}+16\gamma^{4}+3994\gamma^{2}+3677 & h_{54}&=40\gamma^{6}+920\gamma^{4}-751\gamma^{2}+220 \\
h_{55}&=194\gamma^{6}+602\gamma^{4}+265\gamma^{2}+10 & h_{56}&=333\gamma^{6}+1118\gamma^{4}+555\gamma^{2}+24 \\
h_{57}&=300\gamma^{6}+5145\gamma^{4}-322\gamma^{2}+25 & h_{58}&=32\gamma^{6}-336\gamma^{4}-2854\gamma^{2}-967 \\
h_{59}&=544\gamma^{6}+2312\gamma^{4}+1254\gamma^{2}+55 & h_{60}&=48\gamma^{6}+3332\gamma^{4}+7392\gamma^{2}+1723 \\
h_{61}&=125\gamma^{6}+1666\gamma^{4}+3157\gamma^{2}+200 & h_{62}&=480\gamma^{6}+5592\gamma^{4}+4534\gamma^{2}+269 \\
h_{63}&=400\gamma^{6}+7952\gamma^{4}-6379\gamma^{2}+1888 & h_{64}&=2664\gamma^{6}+8992\gamma^{4}+4533\gamma^{2}+201 \\
h_{65}&=450\gamma^{6}+1169\gamma^{4}+3638\gamma^{2}-1792 & h_{66}&=7292\gamma^{6}+19644\gamma^{4}+8141\gamma^{2}+310 \\
h_{67}&=7292\gamma^{6}+19484\gamma^{4}+7905\gamma^{2}+288 & h_{68}&=308\gamma^{6}+6020\gamma^{4}+12803\gamma^{2}+3369 \\
h_{69}&=1000\gamma^{6}+21392\gamma^{4}+41080\gamma^{2}+9353 & h_{70}&=5424\gamma^{6}+80896\gamma^{4}+129626\gamma^{2}+25869
\end{flalign*}
\vskip -25pt
\begin{flalign*}
h_{71}&=8\gamma^{8}+4\gamma^{6}-14\gamma^{4}+1109\gamma^{2}+1413 && \\
h_{72}&=2\gamma^{8}-53\gamma^{6}-1872\gamma^{4}-3415\gamma^{2}-752 && \\
h_{73}&=16\gamma^{8}-52\gamma^{6}+392\gamma^{4}+4733\gamma^{2}+2006 && \\
h_{74}&=32\gamma^{8}-104\gamma^{6}+804\gamma^{4}+9426\gamma^{2}+4047 && \\
h_{75}&=4\gamma^{8}-150\gamma^{6}-5392\gamma^{4}-8837\gamma^{2}-1690 && \\
h_{76}&=32\gamma^{8}+1072\gamma^{6}+3966\gamma^{4}+2313\gamma^{2}+117 && \\
h_{77}&=96\gamma^{8}+3824\gamma^{6}+13228\gamma^{4}+6826\gamma^{2}+301 && \\
h_{78}&=16\gamma^{8}-424\gamma^{6}-14982\gamma^{4}-27575\gamma^{2}-6205 && \\
h_{79}&=240\gamma^{8}+5080\gamma^{6}+15430\gamma^{4}+8197\gamma^{2}+383 && \\
h_{80}&=48\gamma^{8}-472\gamma^{6}+22148\gamma^{4}+105548\gamma^{2}+35853 && \\
h_{81}&=576\gamma^{8}+15392\gamma^{6}+43968\gamma^{4}+19884\gamma^{2}+785 && \\
h_{82}&=4800\gamma^{8}+77520\gamma^{6}-74888\gamma^{4}+17707\gamma^{2}+1888 && \\
h_{83}&=560\gamma^{8}+56104\gamma^{6}+173244\gamma^{4}-88469\gamma^{2}-6304 && \\
h_{84}&=8608\gamma^{8}+71664\gamma^{6}+102228\gamma^{4}+28495\gamma^{2}+783 && \\
h_{85}&=2000\gamma^{8}+24656\gamma^{6}+25429\gamma^{4}-40162\gamma^{2}+15104 && \\
h_{86}&=61792\gamma^{8}+386288\gamma^{6}+477928\gamma^{4}+124149\gamma^{2}+3308 && \\
h_{87}&=59552\gamma^{8}+358368\gamma^{6}+422960\gamma^{4}+104127\gamma^{2}+2618 && \\
h_{88}&=2224\gamma^{8}+300672\gamma^{6}+1488134\gamma^{4}+1230665\gamma^{2}+154975 && \\
h_{89}&=135104\gamma^{8}+3124064\gamma^{6}+11006544\gamma^{4}+8114154\gamma^{2}+989459 && \\
h_{90}&=32\gamma^{10}+288\gamma^{8}-1864\gamma^{6}+15304\gamma^{4}+166837\gamma^{2}+67518 && \\
h_{91}&=32\gamma^{10}+288\gamma^{8}-1724\gamma^{6}+12336\gamma^{4}+115653\gamma^{2}+39000 && \\
h_{92}&=96\gamma^{10}-128\gamma^{8}+10444\gamma^{6}+132316\gamma^{4}+239623\gamma^{2}+57599 && \\
h_{93}&=2816\gamma^{10}+104768\gamma^{8}+567488\gamma^{6}+694456\gamma^{4}+183432\gamma^{2}+4995 && \\
h_{94}&=704\gamma^{10}-30192\gamma^{8}-1977208\gamma^{6}-7574774\gamma^{4}-5557485\gamma^{2}-663020 && \\
h_{95}&=704\gamma^{10}-29872\gamma^{8}-1891304\gamma^{6}-6931946\gamma^{4}-4820363\gamma^{2}-541594 && \\
h_{96}&=64\gamma^{12}+480\gamma^{10}-3424\gamma^{8}+50640\gamma^{6}+1353038\gamma^{4}+1952851\gamma^{2}+361986 && \\
h_{97}&=64\gamma^{12}+480\gamma^{10}-3704\gamma^{8}+51396\gamma^{6}+1350910\gamma^{4}+1926839\gamma^{2}+345760 &&
\end{flalign*}
\endgroup

\section{Details of the post-Newtonian comparison}
\label{app:pn-details}

The tidal correction to the angle follows from the perturbation of the orbit at fixed incoming energy and angular momentum. For the tidal terms, we take the conserved energy and angular momentum from equations~(53a,b), (55a--c), and (C2a,b) in~\cite{Bernard:2023eul}; for the dynamics without tides, we use equations~(4.1a--c) and (4.2a--c) in~\cite{Bernard:2018ivi}. These quantities are expressed in harmonic center-of-mass coordinates, with accelerations eliminated using the equations of motion. We convert them to our conventions using Appendix~\ref{app:laura-notation}.

For planar motion, $r$ and $\theta$ are relative polar coordinates, and dots denote derivatives with respect to the center-of-mass coordinate time, so that $\dot r^2+r^2\dot\theta^2$ is the squared coordinate speed, distinct from the asymptotic $v^2$. With binding energy $E_{\rm bind}=E-m$ and orbital angular momentum $J>0$, the conservation laws read
\begin{align}
 E_{\rm bind}(r,\dot r^2,\dot\theta)&=m(\Gamma-1),\qquad
 J(r,\dot r^2,\dot\theta)=b_ep_\infty.
 \label{eq:PN-EJ-conservation}
\end{align}
Solving these relations together to first order in the tidal response coefficients yields the corrections to $\dot r^2$ and $\dot\theta$. Obtaining the 5PN correction requires the dynamics without tides through 2PN, since the leading scalar dipole enters at 3PN.
At Newtonian order, these relations reduce to
\begin{align}
 \dot r_{\rm N}^{\,2}&=\frac{2E_{\rm bind}}{\mu}
 +\frac{2Gm(1+s_1s_2)}{r}-\frac{J^2}{\mu^2r^2},\qquad
 \dot\theta_{\rm N}=\frac{J}{\mu r^2}.
 \label{eq:PN-Newtonian-orbit}
\end{align}
For weak scattering, the deflection angle is
\begin{align}
 \chi(E_{\rm bind},J)&= 2\int_{r_t}^{\infty}
 \frac{\dot\theta}{\dot r}\,\rmd r-\pi,
 \label{eq:PN-full-angle}
\end{align}
where $r_t$ is the turning point and $\dot r>0$ on the outgoing branch. Linearizing the absolute value about the orbit without tides yields the factor $\vartheta/|\vartheta|$ in the angle correction.

We integrate the post-Newtonian expansion term by term from the Newtonian turning point, using the finite-part prescription to account for its shift without a separate endpoint term~\cite{Jain:2023vlf}. The required moments are given by
\begin{align}
 &\int_{r_{t,\rm N}}^\infty\frac{\rmd r}{r^p}
 (\dot r_{\rm N}^{\,2})^{k-1/2}
 =\frac12\left(\frac{\mu}{J}\right)^{p-1}
 \left(\frac{2E_{\rm bind}}{\mu}\right)^{p/2+k-1}
 \nonumber\\
 &\qquad\times\Biggl[
 \mathrm B\left(\frac{p-1}{2},k+\frac12\right)
 +\frac{(2k-1)Gm\mu(1+s_1s_2)}{J\sqrt{2E_{\rm bind}/\mu}}
 \mathrm B\left(\frac p2,k-\frac12\right)\Biggr]+\mathcal{O}(G^2).
 \label{eq:PN-finite-part-moments}
\end{align}
Here $p>1$, $k$ is an integer, and $E_{\rm bind}>0$. The Euler beta functions $\mathrm B$ follow by expanding in $G$ and substituting $y=J/(r\sqrt{2\mu E_{\rm bind}})$; analytic continuation supplies the finite part when the endpoint integral diverges. At fixed dimensionful response coefficients, tidal terms start at $G^2$. Consequently, the 3PM comparison requires the dynamics without tides and the radial moments only through $G^1$. For a $G^3$ tidal term, only the $G^0$ parts of the dynamics without tides and the radial moments contribute.

For example, retaining the Newtonian terms and the leading scalar-dipole response yields
\begin{align}
 & E_{\rm bind} =\frac{\mu}{2}(\dot r^2+r^2\dot\theta^2)
 -\frac{Gm_1m_2(1+s_1s_2)}{r}
 -\frac{G^4m_1^2m_2^2}{2r^4}\left(m_1s_2^2\bar{\lambda}_{1,0}^{(1)}
 +m_2s_1^2\bar{\lambda}_{2,0}^{(1)}\right)+\cdots,
\nonumber\\[0.35 em]
 & J=\mu r^2\dot\theta+\cdots.
 \label{eq:PN-LO-source}
\end{align}
The factor $G^4$ includes the normalization of the dimensionless response. The $r^{-4}$ interaction contributes through the $p=6,k=-1$ moment. After division by $(\mu/J)^5(2E_{\rm bind}/\mu)$, this moment reads
\begin{align*}
 -\frac{3\pi}{4}
 -\frac{8Gm\mu(1+s_1s_2)}{J\sqrt{2E_{\rm bind}/\mu}}+\mathcal{O}(G^2).
\end{align*}
The two terms yield the leading scalar-dipole contributions at 2PM and 3PM, respectively. For the full comparison, we express the angle at fixed $b_e$ and $v$ using
\begin{align}
 E_{\rm bind}&=m(\Gamma-1),\qquad
 J=\frac{\mu b_e}{\Gamma}\sqrt{\gamma^2-1}.
 \label{eq:fixed-b-dictionary}
\end{align}
Expanding these relations in the asymptotic relative speed gives
\begin{align}
 \frac{E_{\rm bind}}{\mu}&=\frac{v^2}{2}
 +\frac{3-\nu}{8}v^4
 +\frac{5-3\nu+\nu^2}{16}v^6+\mathcal{O}(v^8),\nonumber\\
 \frac{J}{\mu b_e}&=v\left[1+\frac{1-\nu}{2}v^2
 +\frac{3-5\nu+3\nu^2}{8}v^4+\mathcal{O}(v^6)\right].
 \label{eq:PN-asymptotic-expansions}
\end{align}
These substitutions must be made before the final truncation because lower-order terms generate part of the 4PN and 5PN scalar-dipole contributions. The 5PN result also requires the next-to-next-to-leading dipolar terms in both conserved quantities, including those proportional to the first scalar derivative of the response.

In the post-Minkowskian result, we likewise expand the total-energy factor:
\begin{align}
 \Gamma&=1+\frac{\nu}{2}v^2
 +\frac{3\nu-\nu^2}{8}v^4+\mathcal{O}(v^6).
 \label{eq:PN-Gamma-expansion}
\end{align}

\providecommand{\href}[2]{#2}\begingroup\raggedright\endgroup

\end{document}